\documentclass[11pt, a4paper]{extarticle}
\pdfoutput=1

\usepackage{jheppub}
\usepackage{bm}

\newcommand\al[1]{\begin{align}#1\end{align}}
\newcommand\aln[1]{\begin{align*}#1\end{align*}}
\newcommand\als[1]{\begin{align}\begin{split}#1\end{split}\end{align}}

\def\cA{\mathcal{A}}

\def\cV{\mathcal{V}}

\newcommand{\del}{\partial}
\DeclareMathOperator{\Tr}{Tr}

\preprint{RUP-25-5}

\title{Non-perturbative effects in JT gravity from KdV equations}

\author[a]{Yasuyuki Hatsuda,}
\author[b]{Takaki Matsumoto}
\author[c]{and Kazumi Okuyama}

\affiliation[a]{Department of Physics, Rikkyo University,\\Toshima, Tokyo 171-8501, Japan}
\affiliation[b]{Seikei University,\\3-3-1 Kichijoji-Kitamachi, Musashino-shi, Tokyo, 180-8633, Japan}
\affiliation[c]{Department of Physics, Shinshu University,\\3-1-1 Asahi, Matsumoto 390-8621, Japan}

\emailAdd{yhatsuda@rikkyo.ac.jp}
\emailAdd{takaki-matsumoto@ejs.seikei.ac.jp}
\emailAdd{kazumi@azusa.shinshu-u.ac.jp}

\abstract{It is well-known that the partition function of the Jackiw-Teitelboim (JT) gravity is obtained by an integral transformation of volumes of moduli spaces for Riemann surfaces, also known as the Weil-Petersson volumes. This fact enables us to compute the perturbative genus expansion of the partition function by solving a KdV-type non-linear partial differential equation. In this work, we find that this KdV equation also admits transseries solutions. We give a systematic algorithm to explicitly construct a one-parameter transseries solution to the KdV equation. Our approach is based on general two-dimensional topological gravity, and the results for the JT gravity are easily obtained as a special case. The results in the leading non-perturbative sector perfectly agree with another independent calculation from topological recursions in random matrices.}

\begin{document} 
\maketitle
\flushbottom

\section{Introduction}
The two-dimensional Jackiw-Teitelboim (JT) gravity \cite{Jackiw:1984je,Teitelboim:1983ux} is an interesting 
toy model of quantum gravity and holography. Recently, it has attracted much attention in the context of the black hole information problem 
and the Page curve \cite{Penington:2019kki,Almheiri:2019qdq}.

In the seminal paper \cite{Saad:2019lba},
it was shown that the path integral of the JT gravity is equivalent to
a certain double-scaling limit of a random matrix model.
Therefore, the JT gravity can be thought of as a special case of the matrix model
of 2d gravity studied in the early literature 
\cite{Gross:1989vs,Douglas:1989ve,Brezin:1990rb}.
In fact, it is conjectured that the JT gravity is obtained by the $p\to\infty$
limit of the $(2,p)$ minimal string \cite{unpublished}.
One important difference from the viewpoint of the early literature \cite{Gross:1989vs,Douglas:1989ve,Brezin:1990rb} is that the matrix model of the JT gravity is an example of 
holography involving the ensemble average, where the random matrix $H$
is interpreted as a random Hamiltonian of the boundary theory.
As shown in \cite{Saad:2019lba}, the genus-$g$ contribution of the connected correlator
$\langle\prod_{i=1}^n\Tr e^{-\beta_i H}\rangle_c$ 
is constructed from the so-called ``trumpet'' and the Weil-Petersson volume
$V_{g,n}(\bm{b})$.
This identification of the JT gravity as the random matrix model is based on 
the fact that
the Mirzakhani's recursion relation obeyed by the Weil-Petersson volume
\cite{Mirzakhani:2006fta}
is equivalent to the topological recursion of a random matrix model
\cite{Eynard:2007fi}.

It is known that the Weil-Petersson volume grows factorially at large genus
\cite{Zograf:2008wbe}
and hence the genus expansion of the JT gravity is an asymptotic series
and receives non-perturbative corrections \cite{Saad:2019lba}.
In the context of the matrix model of 2d gravity,
such non-perturbative corrections come from the 
eigenvalue instantons in the matrix integral, 
where some of the eigenvalues are pulled out from the cut
and put on the critical points of the effective potential 
\cite{David:1990sk, David:1992za},
and they correspond to ZZ branes \cite{Zamolodchikov:2001ah} in the minimal string theory 
\cite{Hanada:2004im, Kutasov:2004fg}.
More general treatments of non-perturbative corrections in matrix models are found in \cite{Marino:2007te, Marino:2008ya, Marino:2008vx}.
In \cite{Gregori:2021tvs, Eynard:2023qdr}, non-perturbative corrections
of the genus expansion in the JT gravity has been studied from the matrix model
approach.

In this paper, we study the non-perturbative effects
in the JT gravity by another route.
From the Witten-Kontsevich theorem \cite{Witten:1990hr, Kontsevich:1992ti},
it follows that the free energy of the 2d topological gravity
defines a $\tau$-function of the KdV hierarchy.
As shown in \cite{Mulase:2006baa,Dijkgraaf:2018vnm}, 
the generating function of the Weil-Petersson volume
is obtained from the free energy of the topological gravity
by setting the coupling $\{t_k\}_{k\geq0}$ to a specific value $t_k=\gamma_k$
in \eqref{eq:gamma-k}.
This fact allows us to study not only the genus expansion of the JT gravity \cite{Okuyama:2019xbv} but also the non-perturbative corrections by using the KdV equation.
We construct a transseries solution to the KdV equation.
We find that the instanton action for the general
coupling $\{t_k\}$ of topological gravity is determined by
the critical point of the effective potential, as expected from the
matrix model picture. By setting $t_k=\gamma_k$, we
find a complete agreement with the instanton
corrections to the JT gravity computed in \cite{Gregori:2021tvs, Eynard:2023qdr}.

This paper is organized as follows.
In section \ref{sec:JT}, we briefly review a relation between the JT gravity and the Weil-Petersson volume.
In section \ref{sec:top}, we also review the KdV hierarchy in general 2d topological gravity.
In section \ref{sec:trans}, we construct the transseries solution to the KdV equation for 2d topological gravity. We give a systematic algorithm to do so. We emphasize that the solution is valid for general 2d topological gravity. The JT case is included as a special case.
In section \ref{sec:np}, we concretely reduce the results in the previous section to the JT gravity. We compare them with the large genus behavior of the perturbative solution. This is a kind of resurgent analysis. Our obtained results show a perfect agreement with all the available data in the literature.
In section \ref{sec:BA}, we give a comment on a relation between our approach and the random matrix approach.
Finally we conclude in section \ref{sec:conclusion}
with some discussion on the future problems.
We show some detailed results and computations in the appendices.

\section{JT gravity and Weil-Petersson volume}\label{sec:JT}
In this work we focus on the partition function of the JT gravity with $n$ boundaries. We review a basic framework how to compute it. The partition function has a perturbative expansion in a genus counting parameter $\lambda$ as%
\footnote{In some references on the JT gravity, this expansion is called ``non-perturbative'' because $\lambda \sim O(e^{-1/G_N})$, where $G_N$ is the gravitational constant. In this terminology, what we call non-perturbative corrections here is referred to as ``doubly non-perturbative'' corrections. We follow the standard terminology in random matrices \cite{Saad:2019lba}.}
\als
{
	Z_{\text{JT}}(\beta_1,\ldots,\beta_n)
	=
	\sum^\infty_{g=0}\lambda^{2g+n-2}Z_{g,n}(\beta_1,\ldots,\beta_n).
}
where $Z_{g,n}$ is given by a path integral on a hyperbolic surface of genus $g$ with the $n$ boundaries,
\als
{
	\label{genus g contribution to JT}
	Z_{g,n}(\beta_1,\ldots,\beta_n)
	=
	\int^\infty_0
	\Biggl(
	\prod^n_{i=1}\sqrt{\frac{\gamma}{2\pi\beta_i}} b_idb_i
	\,e^{-\gamma b^2_i/2\beta_i}
	\Biggr)
	V_{g,n}(b_1,\ldots,b_n).
}
In this equation, $V_{g,n}(b_1,\ldots,b_n)$ denotes the Weil-Petersson volume of the moduli space of a hyperbolic surface of genus $g$ with $n$ geodesic boundaries of lengths specified by $(b_1,\ldots,b_n)$. We can see that $Z_{g,n}$ is obtained by an integral transform of $V_{g,n}$. Let us consider a generating function of $V_{g,n}$,
\als
{
	\label{eq:Vn-gene}
	\mathcal{V}_n(b_1,\ldots,b_n)
	:=
	\sum^\infty_{g=0}\lambda^{2g+n-2}V_{g,n}(b_1,\ldots,b_n).
}
One way to compute $V_{g,n}(b_1,\ldots,b_n)$ is by using Mirzakhani's recursion relation \cite{Mirzakhani:2006fta, Mirzakhani:2006eta}.

Let $M_{g,n}(\bm{b})$ be the moduli space of hyperbolic Riemann surfaces of genus $g$ with $n$ geodesic boundaries of lengths $\bm{b}=(b_1,\ldots,b_n)$. This space is of complex dimension $3g-3+n$ and carries a natural symplectic form called the Weil-Petersson form. The Weil-Petersson volume $V_{g,n}(\bm{b})$ is the volume of $M_{g,n}(\bm{b})$ with respect to the volume form induced by the Weil-Petersson form.

According to Mirzakhani's theorems \cite{Mirzakhani:2006fta, Mirzakhani:2006eta}, the volume $V_{g,n}(\bm{b})$ is a symmetric polynomial in $b^2_1,\ldots,b^2_n$ of total degree $3g-3+n$. To give a precise form of this polynomial, let $M_{g,n}$ be the moduli space of Riemann surfaces of genus $g$ with $n$ distinct punctures, and let $\overline{M}_{g,n}$ denote its Deligne-Mumford compactification. Let $L_i$ be a complex line bundle over $\overline{M}_{g,n}$ whose fiber at each moduli point is its cotangent space at the $i$-th puncture. We denote by $\psi_i$ the first Chern class of $L_i$. We also need the first Miller-Morita-Mumford class on $\overline{M}_{g,n}$ and denote it by $\kappa$. Then $V_{g,n}(\bm{b})$ takes the form of
\als
{
	\label{WP volume polynomial}
	V_{g,n}(\bm{b})
	=
	\sum_{\sum^n_{i=1}k_i\leq3g-3+n}
	C_g(\bm{k})\,
	\frac{b^{2k_1}_1}{2^{k_1}k_1!}\cdots\frac{b^{2k_n}_n}{2^{k_n}k_n!}
}
with coefficients
\als
{
	\label{coefficients of WP volume polynomial}
	C_g(\bm{k})
	&=
	\int_{\overline{M}_{g,n}}\psi^{k_1}_1\wedge\cdots\wedge\psi^{k_n}_n
	\wedge\exp 2\pi^2\kappa,
}
where the integral over $\overline{M}_{g,n}$ vanishes unless the total degree of the integrated classes equals its dimension $3g-3+n$. The symmetry of $V_{g,n}(\bm{b})$ in $b^2_1,\ldots,b^2_n$ follows from the fact that the integral (\ref{coefficients of WP volume polynomial}) is symmetric on the indices $k_1,\ldots,k_n$, because the wedge product between the two-dimensional classes $\psi_i$ is commutative, and $\overline{M}_{g,n}$ is invariant under every permutation of the $n$ punctures.

The integral (\ref{coefficients of WP volume polynomial}) is an intersection number of the $\psi$ classes and the $\kappa$ class, up to an overall factor. The intersection numbers involving both the $\psi$ classes and the $\kappa$ class are expressible in terms of them involving the $\psi$ classes only, such as
\als
{
	\label{intersection number}
	\langle\tau_{k_1}\cdots\tau_{k_n}\rangle_g
	:=
	\int_{\overline{M}_{g,n}}\psi^{k_1}_1\wedge\cdots\wedge\psi^{k_n}_n,
}
which vanishes unless the dimension condition $\sum^n_{i=1}k_i=3g-3+n$ holds. The order of the letters in the notation $\langle\tau_{k_1}\cdots\tau_{k_n}\rangle_g$ is irrelevant similar to the symmetry of (\ref{coefficients of WP volume polynomial}). In this notation, we have \cite{Mulase:2006baa, Dijkgraaf:2018vnm}
\als
{
	\label{relation for intersection numbers}
	C_g(\bm{k})
	=
	\langle\tau_{k_1}\cdots\tau_{k_n}\exp\sum_{k\geq0}\gamma_k\tau_k\rangle_g,
}
where $\gamma_0,\gamma_1,\ldots$ are infinitely many constants defined by
\als
{
	\gamma_0=\gamma_1=0,
	\qquad
	\gamma_k
	=
	\frac{(-1)^k}{(k-1)!}\quad  (k\geq2).
\label{eq:gamma-k}
}
The expression (\ref{WP volume polynomial}) with the relation (\ref{relation for intersection numbers}) reduces the study of $V_{g,n}(\bm{b})$ to that of the intersection numbers of the $\psi$ classes.

Let us close this section by giving a concise expression of $V_{g,n}(\bm{b})$ combining (\ref{WP volume polynomial}) and (\ref{relation for intersection numbers}). We introduce the exponential generating function for all possible intersection numbers of the $\psi$ classes associated with genus $g$,
\als
{
	\label{generating function for genus g}
	F_g(\{t_k\})
	:=
	\langle\exp\sum_{k\geq0}t_k\tau_k\rangle_g,
}
where $t_0,t_1,\ldots$ are infinitely many formal variables. We also introduce a differential operator acting on the formal variables by
\als
{
	\label{boundary creation operator}
	\hat{V}(b)
	:=
	\sum_{k\geq0}
	\frac{b^{2k}}{2^kk!}\partial_k,\qquad
	\partial_k=\frac{\partial}{\partial t_k}.
}
We call this operator the boundary creation operator \cite{Moore:1991ir}. By using these objects, we define
\als
{
	\label{generalized Weil-Petersson volume}
	V_{g,n}(\bm{b},\{t_k\})
	:=
	\hat{V}(b_1)\cdots \hat{V}(b_n)F_g(\{t_k\}).
}
Then by combining it with (\ref{WP volume polynomial}) and (\ref{relation for intersection numbers}), the Weil-Petersson volume is given by 
\als{
	V_{g,n}(\bm{b})=V_{g,n}(\bm{b},\{\gamma_k\}),
}
as shown in \cite{Okuyama:2020ncd, Okuyama:2021eju}.
We call $V_{g,n}(\bm{b},\{t_k\})$ the generalized Weil-Petersson volume.
In particular, we note that
\begin{align}
	V_{g,n}(\{t_k\})&:=V_{g,n}(\bm{b}=0,\{t_k\})=\partial_0^{n} F_g(\{t_k\}),\\
	V_{g,n}&:=V_{g,n}(\bm{b}=0)=\partial_0^{n} F_g(\{t_k\})|_{t_k=\gamma_k}.
\end{align}

\section{Topological gravity and KdV equation}\label{sec:top}

Let us collect the generating function (\ref{generating function for genus g}) for the intersection numbers of the $\psi$ classes associated with genus $g$ and construct a formal power series in the genus counting parameter $\lambda$ as
\als
{
	\label{free energy of topological gravity}
	F(\lambda,\{t_k\})
	:=
	\sum^\infty_{g=0}\lambda^{2g-2}F_g(\{t_k\}).
}
This is the ``free energy'' of two-dimensional topological gravity, a field theoretical interpretation of intersection theory of the $\psi$ classes. The formal variables $t_0,t_1,\ldots$ play the role of coupling constants in this theory and characterize its background.
Recall that results on the JT gravity are obtained by the specialization:
\als
{
	\label{eq:JT-parameters}
	t_k=\gamma_k.
}

A remarkable property of two-dimensional topological gravity is that the partition function $e^F$ is a tau function of the KdV hierarchy. This is the statement of the Witten-Kontsevich theorem \cite{Witten:1990hr, Kontsevich:1992ti}. In other words, if we define the ``specific heat'' by $u_g=\partial^2_0F_g$ and its generating function%
\footnote{As will be seen in the next section, the generating function $u(\lambda,\{t_k\})$ admits transseries expansions. Strictly speaking, we should distinguish the perturbative generating function $u^{(0)}$ of $u_g$ from the full transseries $u$. In this section, we abuse the notations for simplicity.}
\als
{
	\label{genus expansion of specific heat}
	u(\lambda,\{t_k\})
	=
	\sum^\infty_{g=0}\lambda^{2g}u_g(\{t_k\})
	:=
	\sum^\infty_{g=0}\lambda^{2g}\partial^2_0F_g(\{t_k\}),
}
then we have an infinite set of differential equations
\als
{
	\label{KdV hieralchy}
	\partial_ku=\partial_0R_{k+1}[u],
}
where $R_k[u]$ denote the Gelfand-Dikii differential polynomials of $u$. One can determine these polynomials through recursion relations
\als
{
	(2k+1)\partial_0R_{k+1}
	=
	2u\partial_0R_k+(\partial_0u)R_k+\frac{\lambda^2}{4}\partial^3_0R_k,
}
with an initial condition $R_0=1$. For instance, $R_1[u]=u$, $R_2[u]=u^2/2+\lambda^2 \partial^2_0 u/12$, and so on. In particular, the equation for $k=1$ is the classical KdV equation,
\als
{
	\label{KdV equation}
	\partial_1u
	=
	u\partial_0u+\frac{\lambda^2}{12}\partial^3_0u.
}
A consequence of the Witten-Kontsevich theorem is that the KdV hierarchy with an additional equation called the string equation,
\als
{
	\label{string equation}
	\partial_0F
	=
	\frac{t^2_0}{2}+\sum^\infty_{n=0}t_{n+1}\partial_nF,
}
determines $F$ uniquely. This property is the reason why we base our study on topological gravity.

Itzykson and Zuber made an ansatz for solving the infinite set of differential equations (\ref{KdV hieralchy}). This ansatz is the essence of Zograf's algorithm, which determines $u$ as the power series solution of the KdV equation.
Following \cite{Itzykson:1992ya}, we first define new variables by
\als
{
	\label{Itzykson-Zuber variable}
	I_k
	:=
	\sum^\infty_{n=0}t_{n+k}\frac{u^n_0}{n!}.
}
where $u_0=\partial^2_0F_0$ is the genus zero contribution to $u$. We call these variables the Itzykson-Zuber variables. The genus zero limit of the KdV hierarchy (\ref{KdV hieralchy}) and the string equation (\ref{string equation}) gives \cite{Itzykson:1992ya}
\als
{
	\label{u0 equation}
	u_0
	=
	I_0,
}
In terms of $I_k$, Itzykson and Zuber showed that the first two contributions to (\ref{free energy of topological gravity}) takes the form of
\als
{
	&F_0
	=
	\frac{I^3_0}{6}-\sum^\infty_{k=0}\frac{I^{k+2}_0}{k+2}\frac{t_k}{k!}
	+\frac{1}{2}\sum^\infty_{k=0}\frac{I^{k+1}_0}{k+1}\sum_{l+m=k}\frac{t_l}{l!}\frac{t_k}{k!},\\
	&F_1
	=
	\frac{1}{24}\log \frac{1}{1-I_1},
}
and made an ansatz saying that\footnote{Their original ansatz says more on the form of $F_g$.} $F_g$ ($g\geq2$) are polynomials in $(1-I_1)^{-1}$ and $I_k$ $(k\geq2)$. See  \cite{Eguchi:1994cx, Zhang:2019hly} for a proof of this ansatz.

The Itzykson-Zuber ansatz indicates that $F_g$ $(g\geq1)$ depends on $t_0$ and $t_1$ only through $u_0$ and $I_1$. This suggests that we treat
\als
{
	\label{change of variables}
	&y:=u_0,\\
	&t:=1-I_1,
}
as new independent variables instead of $t_0$, $t_1$. This change of variables with the Itzykson-Zuber variables is the essence of the algorithm of computing $F_g$, which we will review in the following. In this change of variables, we have \cite{Zograf:2008wbe}
\als
{
	\label{change of derivatives}
	&\partial_0
	=
	\frac{1}{t}(\partial_y-I_2\partial_t),\\
	&\partial_1
	=
	-\partial_t+y\partial_0,
}
where $\partial_y:=\partial/\partial y$ and $\partial_t:=\partial/\partial t$. To get these expressions, we used
\als
{
	\label{derivative of IZ variable wrt tk}
	\partial_kI_n
	=
	\frac{I_{n+1}}{t}\frac{y^k}{k!}+\frac{y^{k-n}}{(k-n)!}\theta[k-n],
}
where $\theta[x]$ is the Heaviside step function defined to be $1$ for $x\geq0$ and to be $0$ for $x<0$. One can derive (\ref{derivative of IZ variable wrt tk}) from (\ref{Itzykson-Zuber variable}) and (\ref{u0 equation}).

Now we show that the KdV equation (\ref{KdV equation}) with the change of variables (\ref{change of variables}) determines $F_g$. Substituting the power series (\ref{genus expansion of specific heat}) into (\ref{KdV equation}), collecting the terms of order $\lambda^{2g}$, we obtain
\begin{align}
	\label{eq:u0-recursion}
	-\frac{1}{t}\partial_t(tu_g)
	=
	\sum^{g-1}_{h=1}u_{g-h}\partial_0u_h + \frac{1}{12}\partial^3_0u_{g-1},
\end{align}
for $g\geq 1$. One can easily solve this recursive equation with the initial condition $u_0=y$. Note that the variables $y$, $t$ and $t_k$ $(k\geq 2)$ are independent now. The solution at each step can contain an arbitrary linear term in $t^{-1}$. However, such a term has to be zero because
\als
{
	u_g
	=
	\partial^2_0F_g
	=
	\langle \tau^2_0 \exp \sum_kt_k\tau_k\rangle_g
}
vanishes for all $g$ when $t_0=t_1=\cdots=0$, due to the dimension condition for the intersection numbers. Thus the KdV equation (\ref{KdV equation}) unambiguously determines $u_g$ as a power series solution in $\lambda$. Once we determine $u_g$, we can unambiguously determine $F_g$ by solving the relation $u_g=\partial^2_0F_g$, based on the ansatz that $F_g$ $(g\geq 2)$ are polynomials in $t^{-1}$ and $I_k$ $(k\geq 2)$.

We can also determine the generalized Weil-Petersson volumes (\ref{generalized Weil-Petersson volume}) in terms of the Itzykson-Zuber variables by using (\ref{derivative of IZ variable wrt tk}). More precisely, if we define
\als
{
	\label{eq:I-function}
	\mathcal{I}_k(b)
	=
	\sum^\infty_{n=0}\frac{b^{2(n+k)}}{2^{n+k}(n+k)!}
	\frac{y^n}{n!},
}
then (\ref{derivative of IZ variable wrt tk}) gives
\al
{
	\label{eq:BCO-formula1}
	&\hat{V}(b)I_n
	=
	\frac{I_{n+1}}{t}\mathcal{I}_0(b)+\mathcal{I}_n(b),\\
	\label{eq:BCO-formula2}
	&\hat{V}(b_1)\mathcal{I}_n(b_2)
	=
	\frac{\mathcal{I}_0(b_1)\mathcal{I}_{n+1}(b_2)}{t}.
}
These formulas are sufficient to compute the action of the boundary creation operator (\ref{boundary creation operator}) on $F_g$ and to determine the generalized Weil-Petersson volumes.

\section{Transseries solution}\label{sec:trans}

We have seen that the KdV equation (\ref{KdV equation}) determines the specific heat $u$ of topological gravity as its power series solution in the genus counting parameter $\lambda$. Regarding $\lambda$ as a perturbation parameter, the KdV equation in our context is a singular perturbation of non-linear differential equation because $\lambda$ multiplies the highest derivative. Therefore we expect that the KdV equation has a more general solution called a transseries solution. This solution is a formal power series in two ``independent'' parameters: $\lambda$ and an exponentially small factor, such as $\exp [O(1/\lambda)]$. The latter amounts to the order of nonperturbative effects contributing to topological gravity.

To find a transseries solution of the KdV equation, we take a one-parameter transseries ansatz depending on a free parameter $\sigma$,
\als
{
	\label{one-parameter transseries ansatz}
	u(\sigma,\lambda,\{t_k\})
	:=
	\sum^\infty_{\ell=0}\sigma^\ell
	u^{(\ell)}(\lambda,\{t_k\}),
}
where $u^{(0)}$ is the original perturbative series 
\als
{
u^{(0)}(\lambda,\{t_k\})=\sum_{g=0}^\infty \lambda^{2g} u_g(\{t_k\})),
}
and the other $u^{(\ell)}$ $(\ell\geq1)$ are exponentially small contributions. We call $u^{(\ell)}$ the $\ell$-instanton sector for convenience. The KdV equation has a more general transseries solution, a multi-parameter transseries solution depending on two or more free parameters, but it is beyond the scope of this paper. Plugging (\ref{one-parameter transseries ansatz}) into (\ref{KdV equation}) and collecting the terms of order $\sigma^\ell$, we find that $u^{(\ell)}$ obeying the following inhomogeneous differential equation,
\als
{
	\label{equation for l-instanton sector}
	\partial_1 u^{(\ell)}
	=
	\sum^\ell_{m=0}
	u^{(\ell-m)}\partial_0u^{(m)}
	+
	\frac{\lambda^2}{12}\partial^3_0u^{(\ell)}.
}
We would like to solve these equations recursively.
A typical ansatz for $u^{(\ell)}$ is of the form
\als
{
	\label{ansatz for l-instanton sector}
	u^{(\ell)}(\lambda,\{t_k\})
	=
	\lambda^{\ell\beta}
	\exp\biggl[\frac{\ell A(\{t_k\})}{\lambda}\biggr]
	\phi^{(\ell)}(\lambda,\{t_k\}),
}
where $\beta$ is a characteristic exponent, $A$ an ``instanton action'', and $\phi^{(\ell)}$ a formal power series of the form
\als
{
	\label{power series of l-instanton sector}
	\phi^{(\ell)}(\lambda,\{t_k\})
	=
	\sum^\infty_{h=0}\lambda^h\phi^{(\ell)}_h(\{t_k\}).
}
Note that the equation (\ref{equation for l-instanton sector}) gives no restrictions on $\beta$.  We will determine $\beta$ so that the transseries solution has a resurgence structure. In this section, we give a precise form of $A$ and an algorithm to determine $\phi^{(\ell)}$.

\subsection{Instanton action}

As a first step, we determine the instanton action $A$. This is done by looking at the $\ell=1$ sector:
\als
{
	\label{equation for u^1}
	\biggl(
	\partial_1-u^{(0)}\partial_0-\partial_0u^{(0)}-\frac{\lambda^2}{12}\partial^3_0
	\biggr)
	u^{(1)}
	=
	0
}
Substituting the ansatz \eqref{ansatz for l-instanton sector} for $u^{(1)}$ into this equation,
the leading contribution yields
\als
{
	\label{equation for instanton action}
	\partial_tA+\frac{1}{12}(\partial_0A)^3=0.
}
We give a solution to this non-linear equation in a very heuristic way.
First, we define
\begin{align}
	V_{\text{eff}}(\xi,\{t_k\})
	&=
	-\frac{2tz^3}{3}+2\sum^\infty_{k=2}\frac{I_k}{(2k+1)!!}z^{2k+1},
\end{align}
where $z:=\sqrt{2(\xi-y)}$ and $\xi$ is a formal variable. Note that this quantity has already been introduced in \cite{Okuyama:2020vrh} by constructing the genus expansion of the Baker-Akhiezer function. We will briefly explain why it is just below. We also introduce a ``critical value'' $\xi_\ast=\xi_\ast(\{t_k\})$ by
\begin{align}
	\label{definition of xi ast}
	0
	=
	\partial_\xi V_{\text{eff}}(\xi,\{t_k\})|_{\xi=\xi_*}
	=
	-2tz_\ast+2\sum^\infty_{k=2}\frac{I_k}{(2k-1)!!}z^{2k-1}_\ast,
\end{align}
where $z_\ast=\sqrt{2(\xi_\ast-y)}$.  It is useful to rewrite it as
\begin{equation}
\begin{aligned}
t=\sum_{k=2}^\infty \frac{I_{k}}{(2k-1)!!}z_\ast^{2k-2}.
\end{aligned}
\label{eq:t-zstar}
\end{equation}
Then (\ref{equation for instanton action}) is solved by
\begin{equation}
\begin{aligned}
	A(\{t_k\})
	=
	V_{\text{eff}}(\xi_\ast,\{t_k\}) 
	&=
	-\frac{2tz_*^3}{3}+2\sum^\infty_{k=2}\frac{I_k}{(2k+1)!!}z_*^{2k+1}\\
	&=-\frac{4}{3}\sum_{k=2}^\infty \frac{k-1}{(2k+1)!!}I_k z_\ast^{2k+1}.	
\end{aligned}
	\label{instanton action}
\end{equation}
In fact, we easily find
\begin{align}
	\begin{split}
	\label{del A}
	\partial_t A
	&=
	\partial_tV_{\text{eff}}|_{\xi=\xi_\ast}+\partial_t\xi_\ast \cdot \partial_\xi V_{\text{eff}}|_{\xi=\xi_\ast}
	=
	\partial_tV_{\text{eff}}|_{\xi=\xi_\ast}
	=
	-\frac{2z^3_\ast}{3},\\
	\partial_0A
	&=
	\partial_0V_{\text{eff}}|_{\xi=\xi_\ast}+\partial_0\xi_\ast\cdot \partial_\xi V_{\text{eff}}|_{\xi=\xi_\ast}
	=
	\partial_0V_{\text{eff}}|_{\xi=\xi_\ast}
	=
	2z_\ast,
	\end{split}
\end{align}
due to the criticality of $\xi_\ast$.

To see a connection with the Baker-Akhiezer function, we consider a Lax pair of the original KdV equation as
\als
{
	&
	P
	=
	\frac{\lambda^2}{2}\partial^2_0+u^{(0)},\\
	&
	B
	=
	\frac{\lambda^2}{3}\partial^3_0+u^{(0)}\partial_0+\frac{1}{2}\partial_0u^{(0)}.
}
Then, the KdV equation is equivalent to the Lax equation $[\partial_1-B,P]=0$. Now let us consider linear auxiliary equations
\al
{
	\label{linear equations}
	&
	P\psi
	=
	\xi\psi, \\
	\label{linear equations-2}
	&
	\partial_1\psi
	=
	B\psi.
}
The first equation is regarded as a stationary Schr\"odinger-type equation with potential $u^{(0)}$. The eigenvalue $\xi$ does not depend on $t_0$. Moreover, from the Lax equation, we have  $\partial_1\xi=0$. Therefore $\xi$ is a constant. The second equation describes the ``time dependence'' of the eigenfunction $\psi=\psi(\lambda,\{t_k\};\xi)$.
The eigenfunction $\psi=\psi(\lambda,\{t_k\};\xi)$ is called the Baker-Akhiezer function.

By using the linear auxiliary system (\ref{linear equations}), we find
\als
{
	\biggl(
	\partial_1-u^{(0)}\partial_0-\partial_0u^{(0)}-\frac{\lambda^2}{12}\partial^3_0
	\biggr)
	\psi\partial_0\psi
	=
	0.
}
This means that $\psi\partial_0\psi$ is a solution to \eqref{equation for u^1}.
It is expected that a general solution takes the form of a superposition of $\psi\partial_0\psi$ as
\als
{
	\label{integral representation of 1-instanton sector}
	u^{(1)}
	\sim
	\int^\infty_{-\infty} d\xi\,\psi\partial_0\psi,
}
where we have rescaled $\psi$ so that a weight function over $\xi$ becomes unity. This is possible because the equation for $\ell=1$ is linear and $\xi$ is a constant. (\ref{integral representation of 1-instanton sector}) gives an integral representation of the one-instanton solution. The saddle-point approximation of this integral for the WKB-type ansatz of $\psi$ (see also \eqref{eq:psi-WKB} and \eqref{eq:A-WKB}) implies the solution \eqref{instanton action}.

In Section~\ref{subsec:comparison}, we will also see the fact that the solution \eqref{instanton action} is consistent with the large order behavior of the perturbative solution $u^{(0)}$.

\subsection{One-instanton sector}

Let us proceed to giving an algorithm to determine the formal power series $\phi^{(1)}$ in the one-instanton solution $u^{(1)}$.
For convenience, we consider a WKB-like solution
\als
{
	\label{WKB-like expansion}
	u^{(1)}(\lambda,\{t_k\})
	=
	\lambda^\beta \exp\biggl( \frac{1}{\lambda}\sum^\infty_{h=0}\lambda^hA_h(\{t_k\}) \biggr)
}
rather than \eqref{ansatz for l-instanton sector} itself. The translation between them is straightforward. 
For instance, we have
\als
{
	A_0=A,\qquad e^{A_1}=\phi_0^{(1)},\qquad A_2=\frac{\phi_1^{(1)}}{\phi_0^{(1)}},\qquad \dots.
}
Plugging \eqref{WKB-like expansion} into \eqref{equation for l-instanton sector}, we find
\al
{
	\label{equation for A1}
	DA_1
	&=
	-\frac{1}{t}-\frac{1}{4}\partial_0A_0\partial^2_0A_0,\\
	DA_h
	&=
	-\sum^{[h/2]}_{g=1}u_g\partial_0A_{h-2g}
	-\frac{1}{12}\sum_{\substack{0\leq i+j\leq h\\(i,j)\neq(0,0),(0,h),(h,0)}}
	\partial_0A_i\partial_0A_j\partial_0A_{h-i-j}\nonumber\\
	&
	\label{equation for Ah}
	\quad-\frac{1}{4}\sum^{h-1}_{i=0}\partial_0A_i\partial^2_0A_{h-i-1}
	-\frac{1}{12}\partial^3_0A_{h-2}-c_h,
}
where
\als
{
	c_h=\Biggl\{
	\begin{array}{ll}
	0&(h\ \text{even})\\
	\partial_0u_{(h-1)/2}&(h\ \text{odd})
	\end{array}
}
and
\als
{
	D
	:=
	\partial_t+\frac{1}{4}(\partial_0A_0)^2\partial_0
	=
	\partial_t+z^2_\ast\partial_0.
}
Also $[x]$ denotes the integer part of $x$. We solve these equations recursively.

Let us determine $A_1$ by solving the first equation \eqref{equation for A1}. 
It is easy to show
\als
{
	[\partial_0, \partial_t]=\frac{1}{t}\partial_0.
}
Using this and \eqref{del A}, we can derive a partial differential equation for $z$:
\als
{
	\label{del z}
	\partial_0z_\ast
	=
	-\frac{1}{tz_\ast}-\frac{z^{(1)}_\ast}{z^2_\ast},
}
where $z_\ast^{(n)}:=\partial^n_tz_\ast$.
Therefore
\als
{
	\partial_0^2 A_0=\partial_0(2z_*)=-\frac{2}{tz_\ast}-\frac{2z^{(1)}_\ast}{z^2_\ast}.
}
Then, \eqref{equation for A1} is written as
\als
{
	DA_1
	=
	\frac{z^{(1)}_\ast}{z_\ast}
}
It turns out that a solution to this equation is given by
\als
{
	\label{eq:A1-exact}
	A_1
	=
	\frac{1}{2}\log z^{(1)}_\ast+C_1\biggl(y+\frac{z^2_\ast}{2}\biggr)
}
One can confirm this fact by using
\als
{
	&Dz_\ast^{(1)}=z_\ast^{(2)}-z_\ast^2 \biggl(\partial_t+\frac{1}{t}\biggr)\biggl( \frac{1}{tz_\ast}+\frac{z_\ast^{(1)}}{z_\ast^2} \biggr)
	=\frac{2(z_\ast^{(1)})^2}{z_\ast},\\
	&DC_1\biggl(y+\frac{z^2_\ast}{2}\biggr)=C_1'\biggl(y+\frac{z^2_\ast}{2}\biggr)D\biggl(y+\frac{z^2_\ast}{2}\biggr)=0
}
The function $C_1(y+z^2_\ast/2)$ is a zero-mode of $D$, which cannot be fixed from the differential equation. We need additional information.  In the next section, we will fix it from the large order behavior in the JT gravity case.

We next solve the equations (\ref{equation for Ah}). In the higher order computations, the following formulae are very useful:
\al
{
	\label{del zn}
	\partial_0z^{(n)}_\ast
	=
	\biggl(\partial_t+\frac{1}{t}\biggr)^n\partial_0z_\ast
	=
	-\biggl(\partial^n_t+\frac{n}{t}\partial^{n-1}_t\biggr)
	\biggl(\frac{1}{tz_\ast}+\frac{z^{(1)}_\ast}{z^2_\ast}\biggr).
}
and
\als
{
	Dt
	=
	1-\frac{I_2z^2_\ast}{t},\quad
	DI_k
	=
	-\frac{I_{k+1}z^2_\ast}{t},\quad
	Dz_\ast
	=
	-\frac{z_\ast}{t},
}
and
\als
{
	Dz^{(n)}_\ast
	&=
	z^{(n+1)}_\ast-z^2_\ast\biggl(\partial_t+\frac{1}{t}\biggr)^n
	\biggl(\frac{1}{tz_\ast}+\frac{z^{(1)}_\ast}{z^2_\ast}\biggr)
	\\
	&=
	-\frac{(n-1)z^{(n)}_\ast}{t}+\frac{2(n+1)z^{(1)}_\ast z^{(n)}_\ast}{z_\ast}
	+\cdots.
}
The problem of solving (\ref{equation for Ah}), with the data $u^{(0)}_g$, $A$ and $A_1$ given above, is very similar to that of solving a problem discussed in \cite{Okuyama:2020vrh}. Therefore, we consider the following procedure to solve (\ref{equation for Ah}), following an algorithm proposed in \cite{Okuyama:2020vrh}:
	\begin{enumerate}
	\item
	Express the right-hand side of (\ref{equation for Ah}) as a Laurent polynomial in $t^{-1}$,
	$I_{k}$ $(k\geq2)$, $z_\ast$ and $z^{(n)}_\ast$ $(n\geq1)$ by using (\ref{del z}) and
	(\ref{del zn}). All the terms in this polynomial has non-negative powers of $t^{-1}$,
	$I_k$ $(k\geq2)$ and $z^{(n)}_\ast$ $(n\geq2)$.

	\item
	Let $t^{-m}f(I_k,z_\ast,z^{(n)}_\ast)$ denote the part having the highest power of $t^{-1}$.
	This part can arise only from
	\begin{align}
		\label{highet power part in t}
		D\frac{f(I_k,z_\ast,z^{(n)}_\ast)}{(m-2)t^{m-2}I_2z^2_\ast}.
	\end{align}
Subtract this from the expression obtained by the procedure (i). Then $m$ decreases.

	\item
	Repeat the procedure (ii) up to $m=3$. Then all the terms having power $2$ of $t^{-1}$
	will automatically disappear and its power in the remaining terms will be $1$ or $0$.
	They will also not contain any $I_k$.

	\item
	Let $z^{(l)}$ denote the highest order derivative.
	Let $(z^{(l)}_\ast)^{p_l}\cdots (z^{(2)}_\ast)^{p_2}(z^{(1)}_\ast)^{-p_1}g(z_\ast)$
	denote the term with the highest power of $z^{(l)}_\ast$ that does not contain $t^{-1}$.
	This term can arise only from
	\begin{align}
		\label{highest order derivative}
		D\frac{z_\ast (z^{(l)}_\ast)^{p_l}\cdots (z^{(2)}_\ast)^{p_2}g(z_\ast)}
		{2\{\sum^l_{i=2}p_i(i+1)-(p_1+1)\}(z^{(1)}_\ast)^{p_1+1}}.
	\end{align}
Subtract this from the obtained expression.

	\item
	Repeat the procedure (iv) to $l=2$ and $p_l=1$.
	Then all the terms having power $1$ of $z^{(1)}_\ast$ will automatically disappear.
	Let $(z^{(1)}_\ast)^rh(z_\ast)$ $(r\neq1)$ denote the term with the highest power
	of $z^{(1)}_\ast$ that does not contain $t^{-1}$. This term can arise only from
	\begin{align}
		\label{highest order derivative 2}
		D\frac{z_\ast (z^{(1)}_\ast)^{r-1}h(z_\ast)}{2(r-1)}.
	\end{align}

	\item
	Repeat the procedure (v) until the resulting expression vanishes.

	\end{enumerate}
Note that each $A_h$ has a zero-mode $C_h(y+z^2_\ast/2)$. We observe that when $C_h=0$ for $h\geq 2$ gives the correct answer.

Here we sketch an explicit computation of this algorithm for $A_2$.
The equation for $A_2$ is given by
\begin{equation}
\begin{aligned}
DA_2=-\frac{I_2^2 z_\ast}{6t^4}+\frac{1}{t^3}\biggl( \frac{I_2}{6z_\ast}-\frac{I_3z_\ast}{12} \biggr)+\frac{1}{6t^2 z_\ast^3}
-\frac{z_\ast^{(3)}}{4z_\ast^3 z_\ast^{(1)}}\\
+\frac{(z_\ast^{(2)})^2}{8z_\ast^3(z_\ast^{(1)})^2}+\frac{25z_\ast^{(2)}}{12z_\ast^4}-\frac{10(z_\ast^{(1)})^2}{3z_\ast^5}.
\end{aligned}
\end{equation}
We first eliminate $-I_2^2 z_\ast/(6t^4)$ by using the following identity:
\begin{equation}
\begin{aligned}
D\biggl( \frac{I_2}{12t^2 z_\ast}\biggr)=\frac{I_2^2 z_\ast}{6t^4}+\frac{I_3 z_\ast}{12t^3}-\frac{I_2}{12t^3 z_\ast}.
\end{aligned}
\end{equation}
We thus obtain
\begin{equation}
\begin{aligned}
D\biggl( A_2+\frac{I_2}{12t^2 z_\ast}\biggr)=\frac{I_2}{12t^3 z_\ast}+\frac{1}{6t^2 z_\ast^3}+\cdots-\frac{10(z_\ast^{(1)})^2}{3z_\ast^5}.
\end{aligned}
\end{equation}
Next, we eliminate $I_2/(12t^3 z_\ast)$. This is done by using
\begin{equation}
\begin{aligned}
D\biggl( \frac{1}{12tz_\ast^3} \biggr)=\frac{I_2}{12t^3 z_\ast}+\frac{1}{6t^2 z_\ast^3}.
\end{aligned}
\end{equation}
We find
\begin{equation}
\begin{aligned}
D\biggl( A_2+\frac{I_2}{12t^2 z_\ast}-\frac{1}{12tz_\ast^3}\biggr)=-\frac{z_\ast^{(3)}}{4z_\ast^3 z_\ast^{(1)}}
+\frac{(z_\ast^{(2)})^2}{8z_\ast^3(z_\ast^{(1)})^2}+\frac{25z_\ast^{(2)}}{12z_\ast^4}-\frac{10(z_\ast^{(1)})^2}{3z_\ast^5}.
\end{aligned}
\end{equation}
We have completely removed the terms explicitly depending on $t$. We have just finished step (iii), and proceed to step (iv).
The term $-z_\ast^{(3)}/(4z_\ast^3 z_\ast^{(1)})$ is eliminated by
\begin{equation}
\begin{aligned}
D\biggl(\frac{z_\ast^{(3)}}{16z_\ast^2 (z_\ast^{(1)})^2} \biggr)
=\frac{z_\ast^{(3)}}{4z_\ast^3 z_\ast^{(1)}}+\frac{3(z_\ast^{(2)})^2}{8z_\ast^3(z_\ast^{(1)})^2}-\frac{9z_\ast^{(2)}}{4z_\ast^4}
+\frac{3(z_\ast^{(1)})^2}{2z_\ast^5}\\
+\frac{1}{t}\biggl( \frac{3z_\ast^{(2)}}{4z_\ast^3 z_\ast^{(1)}}-\frac{3z_\ast^{(1)}}{4z_\ast^4} \biggr).
\end{aligned}
\end{equation}
We repeat it until the right hand side becomes zero. We leave the remaining computations for an exercise. 
In Appendix~\ref{S2 solution}, we show explicit forms of $A_2$ and $A_3$.
We implemented this algorithm in \textit{Mathematica}, and obtained $A_h$ up to $h=16$.

\subsection{Multi-instanton sector}

We next determine the formal power series $\phi^{(\ell)}$ in the $\ell$-instanton sector $u^{\ell}$ for $\ell\geq2$. Substituting (\ref{power series of l-instanton sector}) into (\ref{equation for l-instanton sector}) and collecting terms of order $\lambda^{h-1}$, we obtain
\als
{
	\label{eq:Multi-Instanton-Sector}
	-\ell\partial_tA\phi^{(\ell)}_h-\frac{(\ell\partial_0A)^3}{12}\phi^{(\ell)}_h
	&=
	\partial_0A\sum^{\ell-1}_{m=1}
	\sum^h_{g=0}m\phi^{(\ell-m)}_g\phi^{(m)}_{h-g}
	+\sum^{\ell-1}_{m=1}
	\sum^{h-1}_{g=0}\phi^{(\ell-m)}_g\partial_0\phi^{(m)}_{h-g-1}\\
	&
	+\sum^{[(h-1)/2]}_{g=0}\phi^{(\ell)}_{h-2g-1}\partial_0u_g
	+\partial_t\phi^{(\ell)}_{h-1}
	+\frac{\ell^2}{4}\partial_0A\partial_0^2A\phi^{(\ell)}_{h-1}\\
	&
	+\frac{\ell^2}{4}(\partial_0A)^2\partial_0\phi^{(\ell)}_{h-1}
	+\ell\partial_0A\sum^{[h/2]}_{g=1}u_g\phi^{(\ell)}_{h-2g}
	+\frac{\ell}{12}\partial_0^3A\phi^{(\ell)}_{h-2}\\
	&
	+\frac{\ell}{4}\partial_0^2A\partial_0\phi^{(\ell)}_{h-2}
	+\frac{\ell}{4}\partial_0A\partial_0^2\phi^{(\ell)}_{h-2}
	+\sum^{[(h-1)/2]}_{g=1}u_g\partial_0\phi^{(\ell)}_{h-2g-1}\\
	&
	+\frac{1}{12}\partial_0^3\phi^{(\ell)}_{h-3}.
}
where $\phi^{(\ell)}_h=0$ if $h<0$. The derivation of this equation is given in appendix~\ref{app:derivation}.
 Since this is simply a linear \textit{algebraic} equation for $\phi^{(\ell)}_h$, we can uniquely determine $\phi^{(\ell)}_h$ in terms of $A$, $\phi^{(\ell)}_{h'}$  ($h'<h$), and $\phi^{(\ell')}_h$ ($\ell'<\ell$). Thus, once the instanton action and the one-instanton sector are given, one can readily determine $\phi^{(\ell)}$ by using (\ref{eq:Multi-Instanton-Sector}). In the following, we present some explicit results for $\phi^{(\ell)}$.

When $h=0$, (\ref{eq:Multi-Instanton-Sector}) reduces to
\als
{
	\label{recursive equation for h=0}
	\phi^{(\ell)}_0
	=
	-\frac{3}{\ell(\ell^2-1)z^2_\ast}
	\sum^{\ell-1}_{m=1}m\phi^{(\ell-m)}_0\phi^{(m)}_0,
}
where we used (\ref{del A}). Using this equation with
\als
{
	\phi^{(1)}_0=e^{A_1},
}
one can determine $\phi^{(\ell)}_0$ for arbitrary $\ell$ recursively. For instance, we have
\als
{
	&
	\phi^{(2)}_0
	=
	-\frac{1}{2z^2_\ast}e^{2A_1},\\
	&
	\phi^{(3)}_0
	=
	\frac{3}{16z^4_\ast}e^{3A_1},\\
	&
	\phi^{(4)}_0
	=
	-\frac{1}{16z^6_\ast}e^{4A_1}.
}

When $h=1$, we have
\aln
{
	\phi^{(\ell)}_1
	&=
	-\frac{3}{2\ell(\ell^2-1)z_\ast^3}
	\biggl[
	\partial_0A\sum^{\ell-1}_{m=1}\sum^1_{g=0}m\phi^{(\ell-m)}_g\phi^{(m)}_{1-g}
	+\sum^{\ell-1}_{m=1}\phi^{(\ell-m)}_0\partial_0\phi^{(m)}_0\\
	&
	+\phi^{(\ell)}_0\partial_0u_0
	+\partial_t\phi^{(\ell)}_0
	+\frac{\ell^2}{4}\partial_0A\partial^2_0A\phi^{(\ell)}_0
	+\frac{\ell^2}{4}(\partial_0A)^2\partial_0\phi^{(\ell)}_0
	\biggr],
}
or equivalently
\aln
{
	\phi^{(\ell)}_1
	&=
	-\frac{3}{2\ell(\ell^2-1)z_\ast^3}
	\biggl[
	2z_\ast\sum^{\ell-1}_{m=1}\sum^1_{g=0}m\phi^{(\ell-m)}_g\phi^{(m)}_{1-g}
	+\sum^{\ell-1}_{m=1}\phi^{(\ell-m)}_0\partial_0\phi^{(m)}_0\\
	&
	+\frac{1}{t}\phi^{(\ell)}_0
	+\partial_t\phi^{(\ell)}_0
	-\ell^2\biggl(\frac{1}{t}+\frac{z^{(1)}_\ast}{z_\ast}\biggr)\phi^{(\ell)}_0
	+\ell^2z_\ast^2\partial_0\phi^{(\ell)}_0
	\biggr].
}
Using this equation with the previously determined data $\phi^{(\ell)}_0$ and
\als
{
	\phi^{(1)}_1=A_2e^{A_1},
}
one can determine $\phi^{(\ell)}_1$ for arbitrary $\ell$ recursively. For instance, we obtain
\als
{
	&\phi^{(2)}_1
	=
	\biggl[
	\frac{I_2}{12t^2z^3_\ast}
	+\frac{13}{24tz^5_\ast}
	+\frac{31z^{(1)}_\ast}{24z^6_\ast}
	-\frac{11z^{(2)}_\ast}{24z^5_\ast z^{(1)}_\ast}
	-\frac{(z^{(2)}_\ast)^2}{12z^4_\ast(z^{(1)}_\ast)^3}
	+\frac{z^{(3)}_\ast}{16z^4_\ast(z^{(1)}_\ast)^2}
	\biggr]
	e^{2A_1},\\
	&\phi^{(3)}_1
	=
	\biggl[
	-\frac{3I_2}{64t^2z^5_\ast}
	-\frac{29}{64tz^7_\ast}
	-\frac{31z^{(1)}_\ast}{32z^8_\ast}
	+\frac{39z^{(2)}_\ast}{128z^7_\ast z^{(1)}_\ast}
	+\frac{3(z^{(2)}_\ast)^2}{64z^6_\ast(z^{(1)}_\ast)^3}
	-\frac{9z^{(3)}_\ast}{256z^6_\ast(z^{(1)}_\ast)^2}
	\biggr]
	e^{3A_1},\\
	&\phi^{(4)}_1
	=
	\biggl[
	\frac{I_2}{48t^2z^7_\ast}
	+\frac{91}{384tz^9_\ast}
	+\frac{187z^{(1)}_\ast}{384z^{10}_\ast}
	-\frac{7z^{(2)}_\ast}{48z^9_\ast z^{(1)}_\ast}
	-\frac{(z^{(2)}_\ast)^2}{48z^8_\ast(z^{(1)}_\ast)^3}
	+\frac{z^{(3)}_\ast}{64z^8_\ast(z^{(1)}_\ast)^2}
	\biggr]
	e^{4A_1}.
}

When $h=2$, we have
\aln
{
	\phi^{(\ell)}_2
	&=
	-\frac{3}{2\ell(\ell^2-1)z_\ast^3}
	\biggl[
	\partial_0A\sum^{\ell-1}_{m=1}
	\sum^2_{g=0}m\phi^{(\ell-m)}_g\phi^{(m)}_{2-g}
	+\sum^{\ell-1}_{m=1}
	\sum^1_{g=0}\phi^{(\ell-m)}_g\partial_0\phi^{(m)}_{1-g}\\
	&
	+\phi^{(\ell)}_1\partial_0u_0
	+\partial_t\phi^{(\ell)}_1
	+\frac{\ell^2}{4}\partial_0A\partial_0^2A\phi^{(\ell)}_1
	+\frac{\ell^2}{4}(\partial_0A)^2\partial_0\phi^{(\ell)}_1\\
	&
	+\ell\partial_0Au_1\phi^{(\ell)}_0
	+\frac{\ell}{12}\partial_0^3A\phi^{(\ell)}_0
	+\frac{\ell}{4}\partial_0^2A\partial_0\phi^{(\ell)}_0
	+\frac{\ell}{4}\partial_0A\partial_0^2\phi^{(\ell)}_0
	\biggr].
}

\section{Non-perturbative effects in JT gravity}\label{sec:np}
Now we come back to the JT gravity. The JT gravity corresponds to the specialization \eqref{eq:JT-parameters}.
In this case, we have the following analytic form of the Itzykson-Zuber variable:
\begin{equation}
\begin{aligned}
I_k=\sum_{n=0}^\infty \frac{(-1)^{n+k}y^n}{(n+k-1)!n!}=(-1)^k \frac{J_{k-1}(2\sqrt{y})}{(\sqrt{y})^{k-1}}\qquad (k\geq 2),
\end{aligned}
\end{equation}
where $J_n(z)$ is the Bessel function of the first kind.
The independent parameters are now $y$ and $t$.
After  constructing the transseries solution, we further set
\begin{equation}
\begin{aligned}
y=0,\qquad t=1,
\end{aligned}
\end{equation}
to obtain non-perturbative corrections in the JT gravity.
It is quite easy to translate the results in the previous section into this special case. 

\subsection{Comparison with large order behaviors}\label{subsec:comparison}
It is well-known that non-perturbative corrections are related to the large order behavior of the perturbative series by the resurgent analysis. According to \cite{Eynard:2023qdr}, the resurgent asymptotics in the large genus limit of the Weil-Petersson volume $V_{g,n}:=V_{g,n}(\bm{b}=0)$ for $n=2$ in the JT gravity is given by
\begin{equation}
\begin{aligned}
V_{g,2} \sim \frac{\Gamma(2g-\frac{1}{2})}{\pi (-A)^{2g-\frac{1}{2}}}\phi_0^{(1)}\left(1-\frac{A}{2g-\frac{3}{2}}\frac{\phi_1^{(1)}}{\phi_0^{(1)}}+\cdots \right) \qquad ( g \to \infty),
\end{aligned}
\end{equation}
where
\begin{equation}
\begin{aligned}
A=-\frac{\pi}{\sqrt{2}},\qquad \phi_0^{(1)}=\frac{1}{2^{1/4}},\qquad \frac{\phi_1^{(1)}}{\phi_0^{(1)}}=\frac{1}{(\sqrt{2}\pi)^3}\left(\frac{4}{3}-\frac{5\pi^2}{6} \right).
\end{aligned}
\label{eq:LOB-values}
\end{equation}
Recall $V_{g,2}=u_g(y=0, t=1)$.
We have computed the perturbative correction $u_g(y,t)$ up to $g=48$ by using \eqref{eq:u0-recursion}, and have checked it numerically.
Inspired by this result, we infer that the perturbative solution $u_g(y,t)$ itself also behaves as
\begin{equation}
\begin{aligned}
u_g(y,t) \sim \frac{\Gamma(2g-\frac{1}{2})}{\pi (-A(y,t))^{2g-\frac{1}{2}}} \phi_0^{(1)}(y,t)\left(1-\frac{A(y,t)}{2g-\frac{3}{2}}\frac{\phi_1^{(1)}(y,t)}{\phi_0^{(1)}(y,t)}+\cdots\right) \qquad (g \to \infty).
\end{aligned}
\label{eq:LOB-u0}
\end{equation}
If this is true, the instanton action is given by
\begin{equation}
\begin{aligned}
A(y,t)=-\lim_{g \to \infty} \biggl(\frac{4g^2 u_{g-1}^{(0)}(y,t)}{u_g^{(0)}(y,t)}\biggr)^{1/2}.
\end{aligned}
\end{equation}
Also the unknown constant $\beta$ in \eqref{WKB-like expansion} is fixed by
\begin{equation}
\begin{aligned}
\beta=\frac{1}{2}.
\end{aligned}
\end{equation}
Using the perturbative data up to $g=48$, we find the following behaviors for $A$, $\phi_0^{(1)}$ and $\phi_1^{(1)}/\phi_0^{(1)}$ at $y=0$ and around $t=0$:
\begin{equation}
\begin{aligned}
A(0, t)&=-\frac{4\sqrt{3}t^{5/2}}{5}\left(1+\frac{9t}{28}+\frac{149t^2}{1120}+\cdots \right),\\
\phi_0^{(1)}(0,t)&=\frac{3^{1/4}}{\sqrt{2\pi}}t^{-1/4}\left(1+\frac{9t}{40}+\frac{2643 t^2}{22400}+\cdots \right),\\
\frac{\phi_1^{(1)}(0,t)}{\phi_0^{(1)}(0,t)}&=-\frac{5\sqrt{3}}{96}t^{-5/2}\left(1-\frac{39t}{100}-\frac{83t^2}{12000}+\cdots \right).
\end{aligned}
\label{eq:Ay0t}
\end{equation}
These results should be compared with the transseries solution of the KdV equation in the previous section. 

Let us see the instanton action. We already know the exact results \eqref{eq:t-zstar} and \eqref{instanton action}. For $y=0$, we have
\begin{equation}
\begin{aligned}
I_k|_{y=0}=\frac{(-1)^k}{(k-1)!} \qquad (k\geq 2).
\end{aligned}
\end{equation}
Hence we obtain
\begin{equation}
\begin{aligned}
t=\sum_{k=2}^\infty \frac{(-1)^k}{(2k-1)!!(k-1)!}z_\ast^{2k-2}=1-\frac{\sin(\sqrt{2} z_\ast)}{\sqrt{2} z_\ast}
\end{aligned}
\end{equation}
and
\begin{equation}
\begin{aligned}
A(0,t)&=-\frac{4}{3}\sum_{k=2}^\infty \frac{(-1)^k}{(2k+1)!! (k-2)!}z_\ast^{2k+1} \\
&=-\biggl( \frac{1}{\sqrt{2}}-\frac{\sqrt{2}z_\ast^2}{3} \biggr) \sin(\sqrt{2}z_\ast)+z_\ast \cos(\sqrt{2}z_\ast)
\end{aligned}
\end{equation}
Especially, by setting $z_\ast=\pi/\sqrt{2}$, we have $t=1$ and
\begin{equation}
\begin{aligned}
A(0,1)=-\frac{\pi}{\sqrt{2}},
\end{aligned}
\end{equation}
which exactly coincides with \eqref{eq:LOB-values}.

Moreover, if we consider the power series expansion around $z_\ast=0$, we can eliminate $z_\ast$. We find the following expansion:
\begin{equation}
\begin{aligned}
A(0,t)=-\frac{4\sqrt{3}t^{5/2}}{5}\biggl(1+\frac{9t}{28}+\frac{149t^2}{1120}+\frac{15747t^3}{246400}+\frac{12123117 t^4}{358758400}+\cdots\biggr).
\end{aligned}
\end{equation}
This is in agreement with \eqref{eq:Ay0t}!

Let us proceed to checking $A_1$. By setting $y=0$, we obtain
\begin{equation}
\begin{aligned}
z_\ast^{(1)}=\frac{1}{\del_{z_\ast} t}=\biggl(\frac{\sin(\sqrt{2}z_\ast)}{\sqrt{2}z_\ast^2}-\frac{\cos(\sqrt{2}z_\ast)}{z_\ast} \biggr)^{-1}.
\end{aligned}
\end{equation}
In the limit $t \to 0$ (or $z_\ast \to 0$), $A_1$ in \eqref{eq:A1-exact} becomes
\begin{equation}
\begin{aligned}
A_1&=C_1-\frac{1}{2}\log \biggl( \frac{2z_\ast}{3} \biggr)+\frac{z_\ast^2}{10}+\frac{z_\ast^4}{350}+\cdots\\
&=C_1-\frac{1}{4}\log \biggl( \frac{4t}{3} \biggr)+\frac{9t}{40}+\frac{519t^2}{5600}+\cdots.
\end{aligned}
\end{equation}
and
\begin{equation}
\begin{aligned}
\phi_0^{(1)}=e^{A_1}=e^{C_1}\biggl( \frac{3}{4t} \biggr)^{1/4}\biggl( 1+\frac{9t}{40}+\frac{2643t^2}{22400}+\cdots \biggr).
\end{aligned}
\end{equation}
Comparing this result with \eqref{eq:Ay0t}, we obtain
\begin{equation}
\begin{aligned}
C_1=-\frac{1}{2}\log \pi.
\end{aligned}
\end{equation}
We conclude that the solution $A_1$ in the JT gravity is simply given by
\begin{equation}
\begin{aligned}
A_1=\frac{1}{2}\log \bigg( \frac{z_\ast^{(1)}}{\pi} \biggr).
\end{aligned}
\end{equation}

Now we consider our interested limit $(y, t) \to (0,1)$.
At $y=0$, we have
\begin{equation}
\begin{aligned}
\phi_0^{(1)}=\biggl(\frac{\pi \sin(\sqrt{2}z_\ast)}{\sqrt{2}z_\ast^2}-\frac{\pi \cos(\sqrt{2}z_\ast)}{z_\ast} \biggr)^{-1/2}.
\end{aligned}
\end{equation}
Setting $z_\ast=\pi/\sqrt{2}$, we finally obtain
\begin{equation}
\begin{aligned}
\phi_0^{(1)}=\frac{1}{2^{1/4}},
\end{aligned}
\end{equation}
which precisely agrees with \eqref{eq:LOB-values}!

\subsection{One-instanton corrections}
Now we look at non-perturbative corrections to the generating function of the Weil-Petersson volume in \eqref{eq:Vn-gene}.
We first focus on $\bm{b}=(0,\dots,0)$ for simplicity.
At the perturbative level, we know%
\footnote{Here $\cV_n^{(0)}$ and $F^{(0)}$ are perturbative sums of $V_{g,n}$ and $F_g$, respectively (\textit{i.e.}, \eqref{eq:Vn-gene} and \eqref{free energy of topological gravity} in the previously abused notation).}
\begin{equation}
\begin{aligned}
\cV_n^{(0)}=(\lambda \del_0)^n F^{(0)}|_{y=0,t=1}.
\end{aligned}
\end{equation}
For $n \geq 2$, we can also write it as
\begin{equation}
\begin{aligned}
\cV_n^{(0)}=(\lambda\del_0)^{n-2} u^{(0)}|_{y=0,t=1}.
\end{aligned}
\end{equation}
Our claim is that these relations should hold for the full transseries including the non-perturbative corrections:
\begin{align}
\cV_n&=(\lambda\del_0)^n F|_{y=0,t=1}\\
&=(\lambda\del_0)^{n-2} u|_{y=0,t=1} \qquad (n \geq 2).
\end{align}
This means that the non-perturbative corrections to $\cV_n^{(0)}$ and $F^{(0)}$ should be extracted from $u$, the transseries solution to the KdV equation.
This is what we would like to see in this subsection.

In the one-instanton sector, we have
\begin{equation}
\begin{aligned}
\cV_n^{(1)}=(\lambda\del_0)^{n-2} u^{(1)}|_{y=0,t=1}.
\end{aligned}
\end{equation}
Here,
\begin{equation}
\begin{aligned}
\del_0^{k}u^{(1)}&=\lambda^{\beta}\del_0^{k}\exp \biggl( \frac{1}{\lambda} \sum_{h=0}^\infty \lambda^h A_h \biggr)\\
&=\lambda^{\beta-k}\exp \biggl( \frac{1}{\lambda} \sum_{h=0}^\infty \lambda^h A_h \biggr)\\&\quad \times \biggl[ (\del_0 A_0)^k+\lambda \Bigl( k(\del_0A_0)^{k-1}\del_0 A_1
+\frac{k(k-1)}{2}(\del_0 A_0)^{k-2} \del_0^2 A_0 \Bigr)+O(\lambda^2) \biggr] \\
&=\lambda^{\beta-k} e^{\frac{A_0}{\lambda}}e^{A_1}(\del_0 A_0)^k
\biggl[1+\lambda \biggl( A_2+k\frac{\del_0 A_1}{\del_0 A_0}+\frac{k(k-1)}{2}\frac{\del_0^2 A_0}{(\del_0 A_0)^2}\biggr)+O(\lambda^2)\biggr]
\end{aligned}
\label{eq:del0-u1}
\end{equation}
For $(y,t)=(0,1)$, we have
\begin{equation}
\label{eq:JT-evaluation}
\begin{aligned}
A_0&=A=-\frac{\pi}{\sqrt{2}},\qquad \del_0 A_0=\sqrt{2}\pi,\qquad \del_0^2 A_0=-\frac{4\sqrt{2}}{\pi},\\
A_1&=-\frac{1}{4}\log 2,\qquad \del_0 A_1=0, \qquad
A_2=\frac{1}{(\sqrt{2}\pi)^3}\left( \frac{4}{3}-\frac{5\pi^2}{6}\right).
\end{aligned}
\end{equation}
Therefore we finally obtain
\begin{equation}
\begin{aligned}
\cV_n^{(1)}=\lambda^{\beta}e^{-\frac{\pi}{\sqrt{2}\lambda}}\frac{(\sqrt{2}\pi)^{n-2}}{2^{1/4}}
\biggl[ 1+\frac{\lambda}{(\sqrt{2}\pi)^3}\biggl(-\frac{5\pi^2}{6}-4n^2+20n-\frac{68}{3} \biggr)+O(\lambda^2) \biggr].
\end{aligned}
\label{eq:Vn-1inst}
\end{equation}
Note that this result is derived by assuming $n \geq 2$, but the final result is also valid for $n=0,1$.
This fact suggests us that the result \eqref{eq:del0-u1} is analytically continued to the $k < 0$ regime.

To compare this result with the one in \cite{Eynard:2023qdr}, we introduce the string coupling constant by
\begin{equation}
\begin{aligned}
g_s:=\frac{\lambda}{(\sqrt{2}\pi)^3}.
\end{aligned}
\end{equation}
The instanton action is given by
\begin{equation}
\begin{aligned}
\frac{A}{\lambda}=-\frac{1}{4\pi^2 g_s},
\end{aligned}
\end{equation}
which coincides with Eq. (3.16) in \cite{Eynard:2023qdr}. The coefficient of $g_s$ in \eqref{eq:Vn-1inst},
\begin{equation}
\begin{aligned}
-\frac{5\pi^2}{6}-4n^2+20n-\frac{68}{3},
\end{aligned}
\end{equation}
also agrees with Eq.~(4.82) in \cite{Eynard:2023qdr}.

To compute the one-instanton sector of the free energy, it is convenient to solve $(\lambda \del_0)^2 F^{(1)}=u^{(1)}$ directly. By solving this, we obtain
\als
{
	\label{eq:FE-expansion}
	F^{(1)}
	=
	\lambda^\beta
	e^{A_0/\lambda}\frac{e^{A_1}}{(\partial_0A_0)^2}
	\biggl[
	1+\lambda
	\biggl(
	A_2-\frac{2\partial_0A_1}{\partial_0A_0}+\frac{3\partial^2_0A_0}{(\partial_0A_0)^2}
	\biggr)
	+O(\lambda^2)
	\biggr].
}
For $(y,t)=(0,1)$, this expansion becomes
\begin{equation}
\begin{aligned}
F^{(1)}|_{y=0,t=1}=(2\sqrt{2}\pi^3 g_s)^{\beta} \frac{e^{-\frac{1}{4\pi^2 g_s}}}{2^{5/4}\pi^2}\biggl( 1+ \sum_{h=1}^\infty F_h^{(1)} g_s^h \biggr),
\end{aligned}
\end{equation}
where
\begin{equation}
\begin{aligned}
F_1^{(1)}&=-\frac{68}{3}-\frac{5 \pi ^2}{6}  , \\
F_2^{(1)}&=\frac{12104}{9}+\frac{818 \pi ^2}{9}+\frac{241 \pi ^4}{72}, \\
F_3^{(1)}&=-\frac{10171120}{81}-\frac{311672 \pi ^2}{27}-\frac{175879 \pi ^4}{270}-\frac{163513 \pi ^6}{6480}-\frac{29 \pi ^8}{48}.
\end{aligned}
\end{equation}
We have computed $F_h^{(1)}$ up to $h=17$ from the transseries solution to the KdV equation, and the results up to $h=6$ are in perfect agreement with Eqs.~(4.23)-(4.28) in \cite{Eynard:2023qdr}. 
We can do the same computation for the multi-instanton sector. Unfortunately there is no literature for comparison.

Next, we consider the case where $\bm{b}\neq\bm{0}$. We know that
\als
{
	\mathcal{V}^{(0)}_n(\boldsymbol{b})
	=
	\lambda^n\hat V(b_1)\cdots \hat V(b_n)F^{(0)}|_{y=0,t=1}
}
holds at the perturbative level. This relation should also be generalized to the non-perturbative level. In the one-instanton sector, we have
\als
{
	\mathcal{V}^{(1)}_n(\boldsymbol{b})
	=
	\lambda^n\hat V(b_1)\cdots \hat V(b_n)F^{(1)}|_{y=0,t=1}.
}
Substituting the expansion (\ref{eq:FE-expansion}) into the right-hand side of this equation, we obtain
\als
{
	\label{eq:WP-expansion}
	\mathcal{V}^{(1)}_n(\boldsymbol{b})
	&=
	\lambda^\beta e^{A_0/\lambda}\frac{e^{A_1}}{(\partial_0A_0)^2}
	\\
	&\times
	\Biggl[
	\prod^n_{i=1}\{\hat V(b_i)A_0\}
	+\lambda\Biggl(A_2-\frac{2\partial_0A_1}{\partial_0A_0}
	+\frac{3\partial^2_0A_0}{(\partial_0A_0)^2}\Biggl)\prod^n_{i=1}\{\hat V(b_i)A_0\}\\
	&
	+\lambda\sum_{1\leq i<j\leq n}\{\hat V(b_i)\hat V(b_j)A_0\}
	\prod^n_{\substack{k=1\\k\neq i,j}}\{\hat V(b_k)A_0\}\\
	&
	+\lambda\sum^n_{i=1}\Biggl\{
	\hat V(b_i)A_1
	-\frac{2}{\partial_0A_0}\hat V(b_i)\partial_0A_0
	\Biggr\}
	\prod^n_{\substack{k=1\\k\neq i}}\{\hat V(b_k)A_0\}+O(\lambda^2)\Biggr].
}
See Appendix~\ref{app:BCO-actions} for useful expressions of the actions of $\hat V(b)$ appearing in the right-hand side. For $(y,t)=(0,1)$, using (\ref{eq:JT-evaluation}) and (\ref{eq:BCO-action-formula}), we arrive at
\als
{
	\label{eq:WP-expansion-at-JTcoupling}
	\mathcal{V}^{(1)}_n(\bm{b})
	&=
	\lambda^{1/2}e^{-\pi/\sqrt{2}\lambda}\frac{(\sqrt{2}\pi)^{n-2}}{2^{1/4}}
	\\
	&\times
	\biggl[
	\prod^n_{i=1}\mathcal{S}(\tilde b_i)
	-\frac{\lambda}{(\sqrt{2}\pi)^3}\biggl(\frac{5\pi^2}{6}+\frac{68}{3}\biggr)
	\prod^n_{i=1}\mathcal{S}(\tilde b_i)\\
	&
	-\frac{\lambda}{(\sqrt{2}\pi)^3}4
	\sum_{1\leq i<j\leq n}(4\mathcal{C}(\tilde b_i)\mathcal{C}(\tilde b_j)+1)
	\prod^n_{\substack{k=1\\k\neq i,j}}\mathcal{S}(\tilde b_k)\\
	&
	+\frac{\lambda}{(\sqrt{2}\pi)^3}\sum^n_{i=1}
	\biggl(
	-\frac{1}{2}\tilde b^2_i\mathcal{S}(\tilde b_i)+16\mathcal{C}(\tilde b_i)+8
	\biggr)
	\prod^n_{\substack{k=1\\k\neq i}}\mathcal{S}(\tilde b_k)+O(\lambda^2)
	\biggr],
}
where we introduced $\tilde b=\sqrt{2}\pi b$ \cite{Okuyama:2021eju} and defined the two functions
\als
{
	\mathcal{S}(b) = \frac{b}{2}\sinh \frac{b}{2},\quad
	\mathcal{C}(z) = \frac{1}{2}\sinh \frac{b}{2},
}
for comparison with \cite{Eynard:2023qdr}. The coefficients of $g_s$ in (\ref{eq:WP-expansion-at-JTcoupling}) perfectly agree with Eq. (4.64) and Eq. (4.81) in \cite{Eynard:2023qdr}.

\subsection{Multi-instanton corrections}
Now we proceed to higher-instanton sectors.
We can generally express  the contribution of the $\ell$-instanton correction $u^{(\ell)}$ to the Weil-Petersson volume (with $n$ boundaries of length zero) as
\aln
{
	\mathcal{V}^{(\ell)}_n
	=
	(\lambda\partial_0)^{n-2}u^{(\ell)}|_{y=0,t=1},
}
if $n\geq 2$. From the expansion
\aln
{
	u^{(\ell)}
	=
	\lambda^{\ell\beta}e^{\ell A/\lambda}\sum^\infty_{h=0}\phi^{(\ell)}_h\lambda^h,
}
we have
\aln
{
	\partial_0^ku^{(\ell)}
	&=
	\lambda^{\ell\beta-k}e^{\ell A/\lambda}
	\biggl[
	(\ell \partial_0A)^k\phi^{(\ell)}_0
	+\lambda
	\biggl(
	k(\ell \partial_0A)^{k-1}\partial_0\phi^{(\ell)}_0
	+\frac{k(k-1)}{2}(\ell \partial_0A)^{k-2}(\ell \partial^2_0A)\phi^{(\ell)}_0\\
	&\quad
	+(\ell \partial_0A)^k\phi^{(\ell)}_1
	\biggr)
	+O(\lambda^2)
	\biggr]
	\\
	&=
	\lambda^{\ell\beta}e^{\ell A/\lambda}
	(\ell \partial_0A)^k\phi^{(\ell)}_0
	\biggl[
	1+\lambda
	\biggl(
	\frac{\phi^{(\ell)}_1}{\phi^{(\ell)}_0}
	+\frac{k}{\ell}\frac{1}{\partial_0A}\frac{\partial_0\phi^{(\ell)}_0}{\phi^{(\ell)}_0}
	+\frac{k(k-1)}{2\ell}\frac{\partial^2_0A}{(\partial_0A)^2}
	\biggr)
	+O(\lambda^2)
	\biggr].
}
In particular, when $\ell=2$, we have
\aln
{
	\partial_0^ku^{(2)}
	=
	\lambda^{2\beta-k}e^{2A/\lambda}
	(2\partial_0A)^k\phi^{(2)}_0
	\biggl[
	1+\lambda
	\biggl(
	\frac{\phi^{(2)}_1}{\phi^{(2)}_0}
	+\frac{k}{2}\frac{1}{\partial_0A}\frac{\partial_0\phi^{(2)}_0}{\phi^{(2)}_0}
	+\frac{k(k-1)}{4}\frac{\partial^2_0A}{(\partial_0A)^2}
	\biggr)
	+O(\lambda^2)
	\biggr].
}
For $(y,t)=(0,1)$, using
\aln
{
	&
	A = - \frac{\pi}{\sqrt{2}},\qquad
	\partial_0 A_0 = \sqrt{2}\pi,\qquad
	\partial_0^2 A_0 = - \frac{4\sqrt{2}}{\pi},\\
	&
	\phi^{(2)}_0
	=
	-\frac{1}{\sqrt{2}\pi^2},\qquad
	\partial_0\phi^{(2)}_0
	=
	-\frac{4\sqrt{2}}{\pi^4}\\
	&
	\phi^{(2)}_1
	=
	\frac{5}{12}+\frac{23}{6\pi^2},
}
we obtain
\aln
{
	\mathcal{V}^{(2)}_n
	=
	-\lambda^{2\beta}e^{-\sqrt{2}\pi/\lambda}
	\frac{(2\sqrt{2}\pi)^{n-2}}{\sqrt{2}\pi^2}
	\biggl[
	1+\frac{\lambda}{(\sqrt{2}\pi)^3}
	\biggl(
	-\frac{5\pi^2}{3}
	-2n^2+18n
	-\frac{130}{3}
	\biggr)
	+O(\lambda^2)
	\biggr],
}
which should be valid for all $n\geq0$.

We assume that the non-perturbative correction to the free energy in the $\ell$-instanton sector takes the form
\al
{
	F^{(\ell)}
	=
	\lambda^{\ell\beta}
	e^{\ell A/\lambda}
	\sum^\infty_{h=0}\tilde F^{(\ell)}_h\lambda^h.
}
By solving the equation $(\lambda\partial_0)^2F^{(\ell)}=u^{(\ell)}$ using this ansatz, we obtain
\al
{
	&
	\tilde F^{(\ell)}_0 = \frac{1}{(\ell\partial_0A)^2}\phi^{(\ell)}_0,\\
	&
	\tilde F^{(\ell)}_1 = \frac{1}{(\ell\partial_0A)^2}
	(\phi^{(\ell)}_1 - 2\ell\partial_0A\partial_0\tilde F^{(\ell)}_0
	- \ell\partial_0^2A\tilde F^{(\ell)}_0),\\
	&
	\tilde F^{(\ell)}_h = \frac{1}{(\ell\partial_0A)^2}
	(\phi^{(\ell)}_h - 2\ell\partial_0A\partial_0\tilde F^{(\ell)}_{h-1}
	- \ell\partial_0^2A\tilde F^{(\ell)}_{h-1} - \partial_0^2\tilde F^{(\ell)}_{h-2}).
}

In terms of the string coupling constant $g_s$, the coprrection $F^{(\ell)}$ can be written as
\al
{
	F^{(\ell)}
	=
	(2\sqrt{2}\pi^3g_s)^{\ell\beta}
	e^{-\ell /4\pi^2g_s}
	\tilde F^{(\ell)}_0
	\biggl(1+\sum^\infty_{h=1}F^{(\ell)}_hg_s^h\biggr)
}
where
\al
{
	F^{(\ell)}_h
	=
	(\sqrt{2}\pi)^{3h}\frac{\tilde F^{(\ell)}_h}{\tilde F^{(\ell)}_0}.
}
When $\ell=2$ and for $(y,t)=(0,1)$, we obtain
\als
{
	&
	F^{(2)}_1
	=
	- \frac{1}{8\sqrt{2}\pi^4},\\
	&
	F^{(2)}_2
	=
	- \frac{130}{3} - \frac{5\pi^2}{3},\\
	&
	F^{(2)}_3
	=
	\frac{25820}{9} + \frac{1910\pi^2}{9} + \frac{133\pi^4}{18}.
}
It would be very interesting to compare these predictions with other independent calculations.

\section{Baker-Akhiezer function, Christoffel-Darboux kernel and wave function in topological recursion}\label{sec:BA}
In this section, we show a relationship between our approach and random matrix approach in \cite{Eynard:2023qdr}.
Recall that the Baker-Akhiezer function satisfies the Schr\"odinger equation
\begin{equation}
\begin{aligned}
( \hbar^2 \del_0^2+u^{(0)} )\psi(\xi, t_0)=\xi \psi(\xi, t_0),
\end{aligned}
\label{eq:BA-Sch}
\end{equation}
where the potential $u^{(0)}$ is the perturbative solution to the KdV equation. Here we use $\hbar$ rather than $\lambda$ or $g_s$. These are related by
\begin{equation}
\begin{aligned}
\hbar=\frac{\lambda}{\sqrt{2}}=2\pi^3 g_s.
\end{aligned}
\end{equation}
The WKB solution to this Schr\"odinger equation can be constructed by the standard way in quantum mechanics if the potential $u^{(0)}$ is given. We consider the WKB ansatz 
\begin{equation}
\begin{aligned}
\psi=\exp\biggl( \frac{\cA}{\hbar}\biggr),
\end{aligned}
\label{eq:psi-WKB}
\end{equation}
then $v:=\del_0 \cA$ satisfies the so-called Riccati equation
\begin{equation}
\begin{aligned}
v^2+\hbar \del_0 v=\xi-u^{(0)}.
\end{aligned}
\end{equation}
This can be solved perturbatively in $\hbar$:
\begin{equation}
\begin{aligned}
v=\sum_{n=0}^\infty \hbar^n v_n,
\end{aligned}
\end{equation}
because we know the perturbative expansion of the potential
\begin{equation}
\begin{aligned}
u^{(0)}=\sum_{g=0}^\infty (\sqrt{2}\hbar)^{2g} u_g.
\end{aligned}
\end{equation}
At the lowest order, we have
\begin{equation}
\begin{aligned}
v_0^2=\xi-u_0.
\end{aligned}
\end{equation}
Therefore there are two branches $v_0=\pm \sqrt{\xi-u_0}$. These branches of course give two independent solutions to \eqref{eq:BA-Sch}. Here we focus on the branch $v_0=\sqrt{\xi-u_0}$. The other is obtained by a simple replacement, as explained in \cite{Okuyama:2020ncd}.

The semi-classical coefficients $v_n$ are easily obtained by a recursion relation (see Eq.~(3.13) in \cite{Okuyama:2020ncd}).
We find
\begin{equation}
\begin{aligned}
v_0=\zeta,\qquad v_1=\frac{1}{4t\zeta^2},\qquad
v_2=-\frac{5}{32t^2 \zeta^5}-\frac{1}{t^3}\biggl( \frac{I_2}{8\zeta^3}+\frac{I_3}{24\zeta}\biggr)-\frac{I_2^2}{12t^4 \zeta},
\end{aligned}
\end{equation}
where $\zeta=\sqrt{\xi-u_0}=z/\sqrt{2}$. Once we know $v_n$, the semi-classical expansion of $\cA$ is obtained by $\del_0 \cA=v$.
Putting the ansatz
\begin{equation}
\begin{aligned}
\cA=\sum_{n=0}^\infty \hbar^n \cA_n,
\end{aligned}
\label{eq:A-WKB}
\end{equation}
we solve $\del_0 \cA=v$ order by order, and obtain
\begin{equation}
\begin{aligned}
\cA_0&=-\frac{2}{3}t\zeta^3+\sum_{n=1}^\infty \frac{2^{n+1}I_{n+1}}{(2n+3)!!}\zeta^{2n+3},\\
\cA_1&=-\frac{1}{2}\log \zeta-\frac{1}{2}\log(4\pi), \\
\cA_2&=-\frac{5}{48t\zeta^3}-\frac{I_2}{24t^2 \zeta},\\
\cA_3&=\frac{5}{64t^2 \zeta^6}+\frac{1}{t^3}\biggl( \frac{I_2}{16\zeta^4}+\frac{I_3}{48\zeta^2}\biggr)+\frac{I_2^2}{24t^4\zeta^2}.
\end{aligned}
\end{equation}
Note that, as shown in Appendix \ref{app:BA-WKB}, the odd order part of $\cA$ is easily obtained by the even order part of $v$.

We next consider the Christoffel-Darboux kernel, which is related to the BA function by
\begin{equation}
\begin{aligned}
K(\xi, \eta)=\frac{1}{\hbar}\int_{-\infty}^{t_0} dx \, \psi(\xi,x)\psi(\eta,x),
\end{aligned}
\end{equation}
or equivalently
\begin{equation}
\begin{aligned}
\hbar \del_0 K(\xi, \eta)=\psi(\xi, t_0)\psi(\eta,t_0).
\end{aligned}
\end{equation}
The diagonal component is then given by
\begin{equation}
\begin{aligned}
\hbar \del_0 K(\xi, \xi)=\psi(\xi, t_0)^2.
\end{aligned}
\label{eq:K-diag}
\end{equation}
We have also another useful expression \cite{Okuyama:2021eju}:
\begin{equation}
\begin{aligned}
K(\xi, \eta)=\hbar \frac{\del_0 \psi(\xi, t_0) \psi(\eta, t_0)-\psi(\xi, t_0) \del_0 \psi(\eta, t_0)}{\xi-\eta}.
\end{aligned}
\end{equation}
In this expression, the diagonal part is written as
\begin{equation}
\begin{aligned}
K(\xi, \xi)=\hbar \bigl(\psi(\xi, t_0) \del_0\del_\xi \psi(\xi, t_0)-\del_\xi \psi (\xi, t_0)\del_0 \psi(\xi, t_0)\bigr).
\end{aligned}
\end{equation}
Plugging the WKB form \eqref{eq:psi-WKB} into it and recalling $\del_0 \cA=v$, we obtain
\begin{equation}
\begin{aligned}
K(\xi, \xi)=\del_0\del_\xi \cA \exp \biggl( \frac{2\cA}{\hbar} \biggr)=\exp \biggl[ \frac{1}{\hbar}\biggl( 2\cA+\hbar \log \del_\xi v \biggr) \biggr].
\end{aligned}
\end{equation}
This allows us to compute the WKB expansion of $K(\xi, \xi)$ systematically:
\begin{equation}
\begin{aligned}
K(\xi, \xi)=\exp\biggl( \frac{1}{\hbar} \sum_{n=0}^\infty \hbar^n K_n\biggr).
\end{aligned}
\end{equation}
After some computations, we arrive at
\begin{equation}
\begin{aligned}
K_0&=2\cA_0, \\
K_1&=-2\log \zeta-\log(4\pi),\\
K_2&=-\frac{17}{24t\zeta^3}-\frac{I_2}{12t^2\zeta},\\
K_3&=\frac{13}{16t^2 \zeta^6}+\frac{I_2}{2t^3 \zeta^4}+\frac{I_2^2}{6t^4 \zeta^2}+\frac{I_3}{12t^3 \zeta^2}.
\end{aligned}
\end{equation}
We claim that under the parameter identification: $\hbar=2\pi^3 g_s$ and $\xi=\pi^2 x$, a function defined by
\begin{equation}
\begin{aligned}
\Psi(x;g_s):=K(\xi, \xi)|_{y=0,t=1},
\end{aligned}
\label{eq:Psi-TR}
\end{equation}
is the wave function appearing in the topological recursion context in \cite{Eynard:2023qdr}.
In fact, the WKB form of the diagonal CD kernel is given by
\begin{equation}
\begin{aligned}
K(\xi,\xi)|_{y=0,t=1}=\exp\biggl( \frac{1}{g_s} \sum_{n=0}^\infty g_s^n \mathbb{S}_n \biggr)
\end{aligned}
\end{equation}
where
\begin{equation}
\begin{aligned}
\mathbb{S}_2 &= 2\pi^3 K_2|_{y=0,t=1}=-\frac{\pi^2}{6\sqrt{x}}-\frac{17}{12x^{3/2}},\\
\mathbb{S}_3 &= (2\pi^3)^2 K_3|_{y=0,t=1}=\frac{\pi^4}{2x}+\frac{2\pi^2}{x^2}+\frac{13}{4x^3}.
\end{aligned}
\end{equation}
See Eqs.~(4.17) and (4.18) in \cite{Eynard:2023qdr}.
The main advantage of this connection is that we can use the Schr\"odinger equation for the BA function to construct $\Psi(x;g_s)$. This route presents us a quite systematic way to obtain the WKB solution.
Once we know the wave function \eqref{eq:Psi-TR}, one can also compute the one-instanton corrections by using the saddle-point method, explained in \cite{Eynard:2023qdr}.

Note that this wave function solves the so-called quantum curve, a quantization of the spectral curve of the JT gravity:
\begin{equation}
\begin{aligned}
\biggl( g_s^2 \frac{d^2}{dx^2}-Q(x;g_s) \biggr) \Psi(x;g_s)=0,
\end{aligned}
\end{equation}
where
\begin{equation}
\begin{aligned}
Q(x;g_s)&=\frac{\sin^2(2\pi \sqrt{x})}{4\pi^2}+g_s\biggl( \frac{\sin(2\pi \sqrt{x})}{\pi \sqrt{x}}-\frac{\cos(2\pi \sqrt{x})}{2\sqrt{x}} \biggr)\\
&\quad+g_s^2\biggl( \frac{2}{x^2}-\frac{(51+2\pi^2 x)\sin(2\pi \sqrt{x})}{24\pi x^{5/2}} \biggr)+O(g_s^3).
\end{aligned}
\end{equation}
It seems that the quantum curve receives an infinite number of quantum corrections, and it is hard to find its closed form.

\section{Conclusion}\label{sec:conclusion}

In this paper we have studied non-perturbative corrections in the JT gravity
from the transseries solution to the KdV equation.
We explicitly constructed a one-parameter transseries solution to the KdV equation by a systematic algorithm. Our approach is based on general 2d topological gravity, and the results for the JT gravity are easily obtained as a special case 
$t_k=\gamma_k$. The results perfectly agree with another independent calculation from topological recursions in random matrices \cite{Eynard:2023qdr}.

There are many interesting open questions.
Our formalism is valid for multi-instanton sectors, and it is intriguing to compare it with other methods, particularly the matrix model approach.
In this paper, we have only considered a one-parameter transseries solution of the
KdV equation.
For full generality, we need to consider the
multi-parameter transseries solution.
It would be interesting to study the general structure of such a
multi-parameter transseries solution.

Also, in our approach, the Stokes data in the transseries is not determined. We guessed it by the additional resurgent analysis in subsection~\ref{subsec:comparison}.
In this sense, the KdV equation itself seems incomplete to determine the full non-perturbative corrections.
The construction of the complete transseries solution is important to see a ``non-perturbative completion'' of the model.
As discussed in \cite{Saad:2019lba}, the effective potential for the 
JT gravity matrix model is not bounded from below, and it seems a problem to define the model non-perturbatively.\footnote{See \cite{Johnson:2019eik} for a possible non-perturbative
completion of the JT gravity matrix model.}
The complete transseries solution, including negative-tension D-branes in \cite{Marino:2022rpz, Schiappa:2023ned}, may resolve this issue.%
\footnote{We thank R. Schiappa for pointing it out.}

\acknowledgments
We thank Ricardo Schiappa for giving us very helpful comments on the draft.
YH was supported in part by JSPS KAKENHI Grant Nos. 22K03641 and 23K25790.
KO was supported
in part by JSPS Grant-in-Aid for Transformative Research Areas (A) 
``Extreme Universe'' 21H05187 and JSPS KAKENHI 22K03594.
\appendix

\section{More on the one-instanton sector}
We here show the explicit expressions for $A_2$ and $A_3$ in the one-instanton sector.
\begin{align}
	\label{S2 solution}
	&A_2
	=
	-\frac{I_2}{12t^2z_\ast}+\frac{1}{12tz^3_\ast}-\frac{7z^{(1)}_\ast}{24z^4_\ast}
	+\frac{5z^{(2)}_\ast}{24z^3_\ast z^{(1)}_\ast}
	+\frac{(z^{(2)}_\ast)^2}{12z^2_\ast (z^{(1)}_\ast)^3}
	-\frac{z^{(3)}_\ast}{16z^2_\ast (z^{(1)}_\ast)^2}\\
	&A_3
	=
	-\frac{1}{8t^2z^6_\ast}-\frac{I_2}{t^2}\biggl(-\frac{z^{(1)}_\ast}{12z^5_\ast}
	+\frac{z^{(2)}_\ast}{48z^4_\ast z^{(1)}_\ast}\biggr)
	-\frac{1}{t}\biggl(\frac{3z^{(1)}_\ast}{8z^7_\ast}
	-\frac{z^{(2)}_\ast}{16z^6_\ast z^{(1)}_\ast}\biggr)
	+\frac{7(z^{(1)}_\ast)^2}{4z^8_\ast}-\frac{11z^{(2)}_\ast}{8z^7_\ast}\nonumber\\
	&\qquad
	-\frac{5(z^{(2)}_\ast)^2}{16z^6_\ast (z^{(1)}_\ast)^2}
	+\frac{17z^{(3)}_\ast}{48z^6_\ast z^{(1)}_\ast}
	-\frac{(z^{(2)}_\ast)^3}{4z^5_\ast (z^{(1)}_\ast)^4}
	+\frac{7z^{(2)}_\ast z^{(3)}_\ast}{24z^5_\ast (z^{(1)}_\ast)^3}
	-\frac{z^{(4)}_\ast}{16z^5_\ast (z^{(1)}_\ast)^2}
	-\frac{(z^{(2)}_\ast)^4}{8z^4_\ast (z^{(1)}_\ast)^6}\nonumber\\
	&\qquad
	+\frac{3(z^{(2)}_\ast)^2z^{(3)}_\ast}{16z^4_\ast (z^{(1)}_\ast)^5}
	-\frac{(z^{(3)}_\ast)^2}{32z^4_\ast (z^{(1)}_\ast)^4}
	-\frac{z^{(2)}_\ast z^{(4)}_\ast}{24z^4_\ast (z^{(1)}_\ast)^4}
	+\frac{z^{(5)}_\ast}{192z^4_\ast (z^{(1)}_\ast)^3}.
\end{align}


\section{Sequence for the leading term of $\ell$-instanton sector}

We begin with
\als
{
	\phi^{(\ell)}_0
	=
	-\frac{3}{\ell(\ell^2-1)z^2_\ast}
	\sum^{\ell-1}_{m=1}m\phi^{(\ell-m)}_0\phi^{(m)}_0.
}
This equation suggests that $\phi^{(\ell)}_0$ takes the form
\als
{
	\phi^{(\ell)}_0
	=
	(-1)^{\ell-1}a_\ell\frac{e^{\ell A_1}}{z^{2(\ell-1)}_\ast}
}
where $a_\ell$ is a sequence determined by the recurrence relation
\als
{
	\label{recursive equation for a}
	a_\ell
	=
	\frac{3}{\ell(\ell^2-1)}
	\sum^{\ell-1}_{m=1}ma_{\ell-m}a_m,
}
with the initial condition $a_\ell=1$.

We conjecture that the solution to (\ref{recursive equation for a}) is given by
\als
{
	\label{conjecture}
	a_\ell
	=
	\frac{\ell}{2^{2\ell-2}}\times
	\left\{
	\begin{array}{cl}
	1&\text{if $\ell$ is odd}\\[2pt]
	2^{\nu_2(\ell/2)+b_{\ell/2}-1}&\text{if $\ell$ is even}
	\end{array}
	\right.,
}
where $\nu_2(n)$ is the 2-adic valuation of an integer $n$, defined by
\als
{
	\nu_2(n)
	=
	\max \{k=0,1,2,\ldots\mid \text{$2^k$ divides $n$}\},
}
and $b_n$ is a sequence defined by
\als
{
	b_n
	=
	\left\{
	\begin{array}{cl}
	-1-\nu_2(n/4)&\text{if $n\,\mathrm{mod}\,4=0$}\\[2pt]
	1&\text{if $n\,\mathrm{mod}\,4=1$}\\[2pt]
	0&\text{if $n\,\mathrm{mod}\,4=2$}\\[2pt]
	1&\text{if $n\,\mathrm{mod}\,4=3$}
	\end{array}
	\right..
}
If we compute the sequence $a_\ell$ using (\ref{recursive equation for a}) in \textit{Mathematica} and compare the results with (\ref{conjecture}), then we observe a perfect match.

\section{Derivation of equation (\ref{eq:Multi-Instanton-Sector})}\label{app:derivation}
Here we derive \eqref{eq:Multi-Instanton-Sector}.
The $\ell$-instanton sector is expanded as
\aln
{
	u^{(\ell)}
	=
	\left\{
	\begin{array}{ll}
	\displaystyle\sum^\infty_{g=0}u_g\lambda^{2g}&(\ell=0)\\[10pt]
	\displaystyle\lambda^{\ell\beta}e^{\ell A/\lambda}
	\sum^\infty_{h=0}\phi^{(\ell)}_h\lambda^h&(\ell\geq1)
	\end{array}
	\right..
}
Using this, we expand
\als
{
	\label{equation for l-instanton sector}
	\partial_1 u^{(\ell)}
	=
	\sum^\ell_{m=0}
	u^{(\ell-m)}\partial_0u^{(m)}
	+
	\frac{\lambda^2}{12}\partial^3_0u^{(\ell)},
}
and collect terms of order $\lambda^{h-1}$. In the following, we assume that $\ell\geq2$.

The term in the left-hand side of the equation (\ref{equation for l-instanton sector}) is expanded as
\aln
{
	\lambda^{-\ell\beta}e^{-\ell A/\lambda}\partial_1u^{(\ell)}
	&=
	\sum^\infty_{h=0}\lambda^{h-1}\ell\partial_1A\phi^{(\ell)}_h
	+\sum^\infty_{h=0}\lambda^h\partial_1\phi^{(\ell)}_h\\
	&=
	\sum^\infty_{h=0}\lambda^{h-1}\ell\partial_1A\phi^{(\ell)}_h
	+\sum^\infty_{h=1}\lambda^{h-1}\partial_1\phi^{(\ell)}_{h-1}.
}
Next, we write the first term in the right-hand side of (\ref{equation for l-instanton sector}) as
\als
{
	\label{eq:NonLinearTerm-for-MultiInstantonSector}
	\sum^\ell_{m=0}u^{(\ell-m)}\partial_0u^{(m)}
	=
	u^{(\ell)}\partial_0u^{(0)}
	+u^{(0)}\partial_0u^{(\ell)}
	+\sum^{\ell-1}_{m=1}u^{(\ell-m)}\partial_0u^{(m)}.
}
Each term in (\ref{eq:NonLinearTerm-for-MultiInstantonSector}) is expanded as
\aln
{
	\lambda^{-\ell\beta}e^{-\ell A/\lambda}u^{(\ell)}\partial_0u^{(0)}
	&=
	\sum^\infty_{i=0}\sum^\infty_{g=0}\lambda^{i+2g}\phi^{(\ell)}_i\partial_0u_g\\
	&=
	\sum^\infty_{h=1}\lambda^{h-1}\sum^{[(h-1)/2]}_{g=0}\phi^{(\ell)}_{h-2g-1}\partial_0u_g,
}
and
\aln
{
	&\lambda^{-\ell\beta}e^{-\ell A/\lambda}u^{(0)}\partial_0u^{(\ell)}\\
	&\qquad=
	\sum^\infty_{g=0}\lambda^{2g}u_g\sum^\infty_{i=0}
	(\lambda^{i-1}\ell\partial_0A\phi^{(\ell)}_i+\lambda^i\partial_0\phi^{(\ell)}_i)\\
	&\qquad=
	\sum^\infty_{h=0}\lambda^{h-1}
	\ell\partial_0A\sum^{[h/2]}_{g=0}u_g\phi^{(\ell)}_{h-2g}
	+\sum^\infty_{h=1}\lambda^{h-1}
	\sum^{[(h-1)/2]}_{g=0}u_g\partial_0\phi^{(\ell)}_{h-2g-1},\\
	&\qquad=
	\sum^\infty_{h=0}\lambda^{h-1}
	\biggl(
	\ell y\partial_0A\phi^{(\ell)}_h
	+\ell\partial_0A\sum^{[h/2]}_{g=1}u_g\phi^{(\ell)}_{h-2g}
	\biggr)\\
	&\qquad
	+\sum^\infty_{h=1}\lambda^{h-1}
	\biggl(
	y\partial_0\phi^{(\ell)}_{h-1}
	+\sum^{[(h-1)/2]}_{g=1}u_g\partial_0\phi^{(\ell)}_{h-2g-1}
	\biggr),
}
and
\aln
{
	&\lambda^{-\ell\beta}e^{-\ell A/\lambda}
	\sum^{\ell-1}_{m=1}u^{(\ell-m)}\partial_0u^{(m)}\\
	&\qquad=
	\sum^{\ell-1}_{m=1}
	\sum^\infty_{i=0}\sum^\infty_{g=0}
	(\lambda^{i+g-1}m\partial_0A\phi^{(\ell-m)}_g\phi^{(m)}_i
	+\lambda^{i+g}\phi^{(\ell-m)}_g\partial_0\phi^{(m)}_i)\\
	&\qquad=
	\sum^\infty_{h=0}\lambda^{h-1}\partial_0A\sum^{\ell-1}_{m=1}
	\sum^h_{g=0}m\phi^{(\ell-m)}_g\phi^{(m)}_{h-g}
	+\sum^\infty_{h=1}\lambda^{h-1}\sum^{\ell-1}_{m=1}
	\sum^{h-1}_{g=0}\phi^{(\ell-m)}_g\partial_0\phi^{(m)}_{h-g-1}.
}
Finally, the second term in the right-hand side of (\ref{equation for l-instanton sector}) is expanded as
\aln
{
	&\lambda^{-\ell\beta}e^{-\ell A/\lambda}\frac{\lambda^2}{12}\partial_0^3u^{(\ell)}\\
	&\qquad=
	\frac{1}{12}\sum^\infty_{h=0}\lambda^{h+2}
	\biggl[
	\biggl(\frac{\ell\partial_0A}{\lambda}\biggr)^3\phi^{(\ell)}_h
	+3\frac{\ell\partial_0A}{\lambda}\frac{\ell\partial_0^2A}{\lambda}\phi^{(\ell)}_h
	+3\biggl(\frac{\ell\partial_0A}{\lambda}\biggr)^2\partial_0\phi^{(\ell)}_h\\
	&\qquad
	+\frac{\ell\partial_0^3A}{\lambda}\phi^{(\ell)}_h
	+3\frac{\ell\partial_0^2A}{\lambda}\partial_0\phi^{(\ell)}_h
	+3\frac{\ell\partial_0A}{\lambda}\partial_0^2\phi^{(\ell)}_h
	+\partial_0^3\phi^{(\ell)}_h
	\biggr]\\
	&\qquad=
	\sum^\infty_{h=0}\lambda^{h-1}\frac{(\ell\partial_0A)^3}{12}\phi^{(\ell)}_h
	+\sum^\infty_{h=1}\lambda^{h-1}
	\biggl(
	\frac{\ell^2}{4}\partial_0A\partial_0^2A\phi^{(\ell)}_{h-1}
	+\frac{\ell^2}{4}(\partial_0A)^2\partial_0\phi^{(\ell)}_{h-1}
	\biggr)\\
	&\qquad
	+\sum^\infty_{h=2}\lambda^{h-1}
	\biggl(
	\frac{\ell}{12}\partial_0^3A\phi^{(\ell)}_{h-2}
	+\frac{\ell}{4}\partial_0^2A\partial_0\phi^{(\ell)}_{h-2}
	+\frac{\ell}{4}\partial_0A\partial_0^2\phi^{(\ell)}_{h-2}
	\biggr)
	+\sum^\infty_{h=3}\lambda^{h-1}
	\frac{1}{12}\partial_0^3\phi^{(\ell)}_{h-3}.
}
Plugging these expansions into (\ref{equation for l-instanton sector}), we obtain
\aln
{
	\ell\partial_1A\phi^{(\ell)}_h+\partial_1\phi^{(\ell)}_{h-1}
	&=
	\sum^{[(h-1)/2]}_{g=0}\phi^{(\ell)}_{h-2g-1}\partial_0u_g
	+\ell y\partial_0A\phi^{(\ell)}_h
	+\ell\partial_0A\sum^{[h/2]}_{g=1}u_g\phi^{(\ell)}_{h-2g}\\
	&
	+y\partial_0\phi^{(\ell)}_{h-1}
	+\sum^{[(h-1)/2]}_{g=1}u_g\partial_0\phi^{(\ell)}_{h-2g-1}\\
	&
	+\partial_0A\sum^{\ell-1}_{m=1}
	\sum^h_{g=0}m\phi^{(\ell-m)}_g\phi^{(m)}_{h-g}
	+\sum^{\ell-1}_{m=1}
	\sum^{h-1}_{g=0}\phi^{(\ell-m)}_g\partial_0\phi^{(m)}_{h-g-1}\\
	&
	+\frac{(\ell\partial_0A)^3}{12}\phi^{(\ell)}_h
	+\frac{\ell^2}{4}\partial_0A\partial_0^2A\phi^{(\ell)}_{h-1}
	+\frac{\ell^2}{4}(\partial_0A)^2\partial_0\phi^{(\ell)}_{h-1}\\
	&
	+\frac{\ell}{12}\partial_0^3A\phi^{(\ell)}_{h-2}
	+\frac{\ell}{4}\partial_0^2A\partial_0\phi^{(\ell)}_{h-2}
	+\frac{\ell}{4}\partial_0A\partial_0^2\phi^{(\ell)}_{h-2}
	+\frac{1}{12}\partial_0^3\phi^{(\ell)}_{h-3}.
}
This equation is equivalento to
\aln
{
	-\ell\partial_tA\phi^{(\ell)}_h-\frac{(\ell\partial_0A)^3}{12}\phi^{(\ell)}_h
	&=
	\partial_0A\sum^{\ell-1}_{m=1}
	\sum^h_{g=0}m\phi^{(\ell-m)}_g\phi^{(m)}_{h-g}
	+\sum^{\ell-1}_{m=1}
	\sum^{h-1}_{g=0}\phi^{(\ell-m)}_g\partial_0\phi^{(m)}_{h-g-1}\\
	&
	+\sum^{[(h-1)/2]}_{g=0}\phi^{(\ell)}_{h-2g-1}\partial_0u_g
	+\partial_t\phi^{(\ell)}_{h-1}
	+\frac{\ell^2}{4}\partial_0A\partial_0^2A\phi^{(\ell)}_{h-1}\\
	&
	+\frac{\ell^2}{4}(\partial_0A)^2\partial_0\phi^{(\ell)}_{h-1}
	+\ell\partial_0A\sum^{[h/2]}_{g=1}u_g\phi^{(\ell)}_{h-2g}
	+\frac{\ell}{12}\partial_0^3A\phi^{(\ell)}_{h-2}\\
	&
	+\frac{\ell}{4}\partial_0^2A\partial_0\phi^{(\ell)}_{h-2}
	+\frac{\ell}{4}\partial_0A\partial_0^2\phi^{(\ell)}_{h-2}
	+\sum^{[(h-1)/2]}_{g=1}u_g\partial_0\phi^{(\ell)}_{h-2g-1}\\
	&
	+\frac{1}{12}\partial_0^3\phi^{(\ell)}_{h-3}.
}

\section{Formulae for boundary creation operator}
\label{app:BCO-actions}

In this appendix, we present useful formulas that describe how the boundary creation operator $\hat{V}(b)$ acts on several important quantities.

The action of $\hat{V}(b)$ on the instanton action $A$ is given by
\als
{
	\label{eq:BCO-on-A}
	\hat V(b)A
	&=
	\hat V(b)V_{\text{eff}}|_{\xi=\xi_\ast}\\
	&=
	2\sum^\infty_{k=0}\frac{z^{2k+1}_\ast}{(2k+1)!!}\mathcal{I}_k(b).
}
In the first equality, we used the condition $\partial_\xi V_{\text{eff}}|_{\xi=\xi_\ast}=0$. The second equality follows from a straightforward calculation using (\ref{eq:BCO-formula1}), and $\mathcal{I}_k(b)$ is defined by (\ref{eq:I-function}). Combining (\ref{eq:BCO-on-A}) and
\als
{
	\partial_0\mathcal{I}_n(b)
	=
	\frac{\mathcal{I}_{n+1}(b)}{t},
}
we obtain
\als
{
	\label{eq:BCO-on-zast}
	\hat V(b)z_\ast
	&=
	-\frac{\mathcal{I}_0(b)}{tz_\ast}
	-\frac{z^{(1)}_\ast}{z^2_\ast}
	\sum^\infty_{n=0}\frac{2^nn!}{(2n)!}z^{2n}_\ast\mathcal{I}_n(b).
}
Next, using (\ref{eq:BCO-on-zast}) and a commutation relation
\als
{
	\hat{V}(b)\partial_t
	=
	\partial_t\hat{V}(b)
	+
	\frac{\mathcal{I}_0(b)}{t}\partial_0,
}
we find
\als
{
	\label{eq:BCO-on-z(1)ast}
	\hat{V}(b)z^{(1)}_\ast
	&=
	-\Biggl(\frac{z^{(2)}_\ast}{z^2_\ast}-\frac{2(z^{(1)}_\ast)^2}{z^3_\ast}\Biggr)
	\sum^\infty_{n=0}\frac{2^nn!}{(2n)!}z^{2n}_\ast\mathcal{I}_n(b)
	-\frac{(z^{(1)}_\ast)^2}{z^2_\ast}
	\sum^\infty_{n=1}\frac{2^nn!}{(2n-1)!}z^{2n-1}_\ast\mathcal{I}_n(b).
}
Furthermore, incorporating (\ref{eq:BCO-on-A}), (\ref{eq:BCO-on-zast}) and (\ref{eq:BCO-on-z(1)ast}), we arrive at
\als
{
	\hat{V}(b_1)\hat{V}(b_2)A
	=
	-\frac{2\mathcal{I}_0(b_1)\mathcal{I}_0(b_2)}{tz_\ast}
	-\frac{2z^{(1)}_\ast}{z^2_\ast}
	\Biggl(\sum^\infty_{n=0}\frac{2^nn!}{(2n)!}z^{2n}_\ast\mathcal{I}_n(b_1)\Biggr)
	\Biggl(\sum^\infty_{m=0}\frac{2^mm!}{(2m)!}z^{2m}_\ast\mathcal{I}_m(b_2)\Biggr).
}
In particular, for $(y,t)=(0,1)$, the above expressions simplify to
\als
{
	\label{eq:BCO-action-formula}
	&
	\hat{V}(b)A
	=
	\frac{2}{b_i}\sinh bz_\ast,\\
	&
	\hat{V}(b)z_\ast
	=
	-\frac{1}{z_\ast}-\frac{z^{(1)}_\ast}{z^2_\ast}\cosh bz_\ast,\\
	&
	\hat{V}(b)z^{(1)}_\ast
	=
	-\Biggl(\frac{z^{(2)}_\ast}{z^2_\ast}-\frac{2(z^{(1)}_\ast)^2}{z^3_\ast}\Biggr)\cosh bz_\ast
	-\frac{(z^{(1)}_\ast)^2}{z^2_\ast}b\sinh bz_\ast,\\
	&
	\hat{V}(b_1)\hat{V}(b_2)A
	=
	-\frac{2}{z_\ast}
	-
	\frac{2z^{(1)}_\ast}{z^2_\ast}
	\cosh b_1z_\ast\cosh b_2z_\ast.
}


\section{Odd part in WKB expansion of Baker-Akhiezer function}\label{app:BA-WKB}
We see that the odd order part of $\cA$-function in \eqref{eq:psi-WKB} is easily computed from the even order part of $v$.
To see it, we divide $v$ as
\begin{equation}
\begin{aligned}
v=v_\text{even}+v_\text{odd},\qquad
v_\text{even}=\sum_{k=0}^\infty \hbar^{2k} v_{2k},\qquad
v_\text{odd}=\sum_{k=0}^\infty \hbar^{2k+1} v_{2k+1}
\end{aligned}
\end{equation}
Substituting it into the Riccati equation, we obtain
\begin{equation}
\begin{aligned}
v_\text{even}^2+v_\text{odd}^2+2v_\text{even}v_\text{odd}+\hbar(\del_0 v_\text{even}+\del_0 v_\text{odd})=\xi-u^{(0)}.
\end{aligned}
\end{equation}
By looking at the even/odd order structure, this equation can be split into the following two equations:
\begin{equation}
\begin{aligned}
v_\text{even}^2+v_\text{odd}^2+\hbar \del_0 v_\text{odd}&=\xi-u^{(0)}, \\
2v_\text{even}v_\text{odd}+\hbar \del_0 v_\text{even}&=0.
\end{aligned}
\end{equation}
From the second equation, we get
\begin{equation}
\begin{aligned}
v_\text{odd}=-\frac{\hbar}{2} \frac{\del_0 v_\text{even}}{v_\text{even}}=-\frac{\hbar}{2}\del_0 \log v_\text{even}.
\end{aligned}
\end{equation}
Therefore the odd order part of $\cA$ is given by
\begin{equation}
\begin{aligned}
\cA_\text{odd}=-\frac{\hbar}{2}\log v_\text{even}.
\end{aligned}
\end{equation}

\bibliography{references}
\bibliographystyle{utphys}

\end{document}